\documentclass[aps,prd,amsmath,amsfonts,amssymb,eqsecnum,nofootinbib,
  floatfix,secnumarabic,preprintnumbers,superscriptaddress,nobalancelastpage,
  onecolumn,notitlepage,biblatex]{revtex4-1}
\newcommand{\cosec}{\operatorname{cosec}}
\usepackage{epsfig}
\usepackage{graphicx}
 \usepackage{placeins}
\usepackage{mathrsfs}
\usepackage{amssymb}
\usepackage{float}
\usepackage{verbatim}
\usepackage{amsmath}
\usepackage[toc,page]{appendix}
\usepackage[colorlinks=true,linkcolor=blue]{hyperref}

\expandafter\ifx\csname package@font\endcsname\relax\else
 \expandafter\expandafter
 \expandafter\usepackage
 \expandafter\expandafter
 \expandafter{\csname package@font\endcsname}%
\fi

\begin{document}

\title{Multicomponent Scalar Dark Matter with an Extended Gauge Sector}
\author{Baradhwaj Coleppa}
\email{baradhwaj@iitgn.ac.in}
\affiliation{Indian Institute of Technology Gandhinagar, Gandhinagar 382 055, India}
\author{Kousik Loho}
\email{kousik.loho@iitgn.ac.in}
\affiliation{Indian Institute of Technology Gandhinagar, Gandhinagar 382 055, India}
\author{Agnivo Sarkar}
\email{agnivosarkar@hri.res.in}
\affiliation{Regional Centre for Accelerator-based Particle Physics, Harish-Chandra Research Institute, HBNI, Chhatnag Road, Jhusi, Prayagraj (Allahabad) 211019, India}

\begin{abstract}
We consider an extension of the Standard Model of particle physics with an additional $SU(2)$ gauge sector along with an additional scalar bidoublet and a non-linear sigma field. The neutral components of the bidoublet serve as dark matter candidates by virtue of the bidoublet being odd under a $Z_2$ symmetry. Generic beyond Standard Model constraints like vacuum stability, invisible decay of higgs, Higgs alignment limit and collider constraints on heavy gauge bosons restrict the parameter space of this model. In this multicomponent dark matter scenario, we investigate the interplay between the annihilation and co-annihilation channels originating from the new gauge sector as those contribute to the relic abundance. We also inspect the direct detection constraints on scattering cross-sections of the dark matter particles with the detector nucleons and present our observations.
\end{abstract}
\begin{flushright}
\small{HRI-RECAPP-2023-06}
\end{flushright}
\maketitle

\section{Introduction}
\label{sec:intro}
The Standard Model (SM) of particle physics has been experimentally proven to be a very successful theory over the last few decades. However, there are a few limitations of the SM which motivate physicists to look for Beyond the Standard Model (BSM) scenarios. One such very important drawback with the SM is that it can not explain the particulate nature of the Dark Matter (DM). DM consists of more than one fourth of the energy budget of our universe \cite{Planck:2018vyg,WMAP:2012nax}. In the quest of solving the DM problem, one very popular class of models are those that incorporate a Weakly Interacting Massive Particle (WIMP) in their spectrum \cite{Bertone:2004pz,Bergstrom:2009ib,Arcadi:2017kky}. In a typical WIMP model, there is an extended sector in addition to the SM particles which is weakly coupled to the SM and protected by a stabilising symmetry. Various scalar extensions including singlet extensions \cite{McDonald:1993ex,Burgess:2000yq,Guo:2010hq,Bandyopadhyay:2010cc,He:2008qm,He:2009yd,Cline:2013gha}, two-singlet extensions \cite{Abada:2011qb,Abada:2012hf,Arhrib:2018eex,Hamada:2020wjh,Modak:2013jya,Maniatis:2020ois,Bhattacharya:2017fid,Maity:2019hre,DiazSaez:2021pfw,Basak:2021tnj}, inert 2HDM \cite{LopezHonorez:2006gr,Belyaev:2016lok}, 2HDM$+$scalar singlet \cite{Aoki:2009pf,Bhattacharya:2019fgs}, 2HDM with gauge extension \cite{Rojas-Abatte:2017hqm}, and also various gauge extensions mostly of different kinds of additional $U(1)$ \cite{Bhat:2019yqo,Nam:2020byw,Okada:2010wd,Basak:2013cga,Okada:2019sbb,Biswas:2016yan,Bandyopadhyay:2017bgh} with an extended scalar and/or fermion sector have been studied in this context.

In this work, we extend the gauge sector of SM with an additional $SU(2)$ and also the scalar sector with an additional scalar bidoublet and a non-linear sigma field. The SM-like scalar doublet and the sigma field contribute to the spontaneous symmetry breaking whereas the scalar bidoublet provides the DM candidates protected by an imposed $Z_2$ symmetry. An extended gauge sector introduces co-annihilation channels for the relic calculation as well as modifies the direct detection constraints - the main goal of the present work is to study in this specific model context how these phenomenological considerations affect the parameter space of the model.

The paper is organised as follows: in Sec.~\ref{sec:model} we introduce the model. In Sec.~\ref{sec:constraints} we discuss the model constraints arising from vacuum stability, invisible decay of the Higgs particle, Higgs alignment, and collider searches for heavy gauge bosons. In Sec.~\ref{sec:dmpheno}, we investigate the DM phenomenology motivating the DM candidates and calculating the relic abundance and the direct detection constraints before finally presenting our conclusions in Sec.~\ref{sec:conclusion}. All the relevant couplings are organised in Appendix~\ref{ap:coup} after a few comments on perturbativity in Appendix~\ref{ap:per}. A brief note on the loop level calculations of Higgs decay to two photons is provided in Appendix~\ref{ap:h_gamma_gamma} with appropriate loops relevant to the model under study. The Boltzmann equations and Feynman diagrams relevant for the relic are given in Appendix~\ref{ap:beq}.


\section{The Model}
\label{sec:model}
We consider an $SU(2)$ extension of the gauge sector of the SM. Furthermore, the scalar sector is extended with an additional scalar bi-doublet odd under a $Z_2$ symmetry and a non-linear sigma field (for details see Tab.~\ref{tab:transformation}).
\begin{table}[h!]
\begin{center}
\begin{tabular}{|c|c|}
\hline
Fields& $Z_2$ charges\\
\hline
The Fermionic Sector & +\\
\hline
$\Phi_1$ & +\\
\hline
$\Sigma$& +\\
\hline
$\Phi_2$ (and thus its components $\phi_0,\phi_3,\phi^\pm$)& -\\
\hline
\end{tabular}
\caption{$Z_2$ charges of the states.}
\label{tab:transformation}
\end{center}
\end{table}
The gauge sector can now be written as $SU(2)_0\times SU(2)_1\times U(1)_2$ (where the subscripts help us to keep track of different gauge charges). We will see that three out of the four degrees of freedom of the scalar doublet and three degrees of freedom of the sigma field are eaten up by the six gauge bosons of the gauge sector, thus giving rise to  massive gauge bosons, while the photon remains massless due to the residual $U(1)_{EM}$ symmetry. The electrically neutral components of the scalar bidoublet protected by a $Z_2$ symmetry furnish the dark matter candidates of the universe. The various scalar fields can be conveniently expressed in the following form

\begin{equation}
\Phi_{1} = 
\begin{pmatrix}
G^+\\
\frac{v_1 + h+iG_0 }{\sqrt{2}} 
\end{pmatrix};
\qquad 
\Phi_{2} = \frac{1}{2}
\begin{pmatrix}
\phi_{0} + i\phi_{3} &   i(\phi_1-i\phi_2) \\
i(\phi_1+i\phi_2)      &    \phi_{0} -i\phi_{3}
\end{pmatrix};
\qquad
\Sigma=\exp{\frac{i\sigma^a\pi^a}{F}}.
\end{equation}
After a few redefinitions, the latter two can be recast as

\begin{equation}
\Phi_{2} = \frac{1}{2}
\begin{pmatrix}
\phi_{0} +i \phi_{3} &   i\sqrt{2}\phi^+ \\
i\sqrt{2}\phi^-     &    \phi_{0} -i \phi_{3}
\end{pmatrix};
\qquad
\Sigma=
\begin{pmatrix}
1+\frac{i}{F}\pi^3 &   \frac{i\sqrt{2}}{F}\pi^+ \\
\frac{i\sqrt{2}}{F}\pi^-     &    1-\frac{i}{F}\pi^3 
\end{pmatrix},
\end{equation}
where $\phi^\pm=\frac{1}{\sqrt{2}}(\phi_1\mp i\phi_2)$ and similarly for the $\pi$'s.\\
The scalars transform under these gauge symmetries in the following manner:

\begin{equation}
\Phi_2\rightarrow U_0^\dagger\Phi_2U_1,\qquad \Sigma\rightarrow U_1^\dagger\Sigma U_0,\qquad \Phi_1\rightarrow U_1\Phi_1,U_2\Phi_1.
\end{equation}
Thus the $\Phi_2$ and $\Sigma$ transform as bidoublets under the $SU(2)_0\times SU(2)_1 $whereas the $\Phi_1$ is a doublet under $SU(2)_1$. The $U_i$ signify the unitary transformation matrices corresponding to the gauge symmetry indexed $i$ and the covariant derivatives accordingly take the form

\begin{align*}
D_{\mu}\Phi_{1} & = \partial_{\mu}\Phi_{1} - \frac{ig_{1}}{2}W^{a}_{1\mu}\sigma^{a}\Phi_{1} - \frac{ig_{2}}{2}B_{2\mu}\Phi_{1} \\
D_{\mu}\Phi_{2} & = \partial_{\mu}\Phi_{2} + \frac{ig_{0}}{2}W^{a}_{0\mu}\sigma^{a}\Phi_{2} -\frac{ig_{1}}{2}\Phi_{2}W^{a}_{1\mu}\sigma^{a} \\
D_{\mu}\Sigma & = \partial_{\mu}\Sigma +\frac{ig_{1}}{2} W^{a}_{1\mu}\sigma^{a}\Sigma
- \frac{ig_{0}}{2}\Sigma W^{a}_{0\mu}\sigma^{a}.
 \end{align*}
With these definitions in place, the kinetic energy term of the scalar sector can be written as
\begin{equation}
\mathcal{L} \supset (D^{\mu}\Phi_1)^\dagger (D_\mu\Phi_1) + \textrm{Tr}\,[(D^\mu\Phi_2)^\dagger (D_\mu\Phi_2)] + \frac{F^2}{4}\textrm{Tr}[(D^\mu\Sigma)^\dagger(D_\mu\Sigma)].
\end{equation}
The $SU(2)_0\times SU(2)_1\times U(1)_2$ symmetry is broken down to $SU(2)\times U(1)_2$ as $\Sigma$ develops a vacuum expectation value $F$. We identify this gauge group with $SU(2)_L\times U(1)_Y$ of the SM. Then this symmetry is further broken down to $U(1)_{EM}$ as the $\Phi_1$ develops a vacuum expectation value $v$ (= 246 GeV). After symmetry breaking, collecting the relevant terms, the neutral gauge boson mass matrix can be written as
\begin{equation}
\text{M}^{2}_{N} =\frac{1}{8}
\begin{pmatrix}
g_0^2F^2&-g_0g_1F^2&0\\
-g_0g_1F^2&g_1^2(F^2+v^2)&-g_1g_2v^2\\
0&-g_1g_2v^2&g_2^2v^2
\end{pmatrix}
=\frac{g_0^2F^2}{8}
\begin{pmatrix}
1&-x&0\\
-x&x^2(1+r^2)&-x^2tr^2\\
0&-x^2tr^2&x^2t^2r^2
\end{pmatrix},
\end{equation} where $\frac{g_1}{g_0}= x\ll1$, $\frac{g_2}{g_1}=t$, and $\frac{v}{F}=r$.\\
This matrix can be diagonlized perturbatively in the small parameter $x$. The eigenvalues corresponding to the eigenstates of the photon, $Z$ and $Z^\prime$ bosons are given by
\begin{equation}
m_\gamma^2=0;\qquad m_Z^2=\frac{g_0^2F^2r^2x^2}{4}(1+t^2-x^2);\qquad m_{Z^\prime}^2=\frac{g_0^2F^2}{4}(1+x^2),
\label{eq:neutral_boson_mass}
\end{equation}
and the mass eigenstates are (writing $t=\frac{s}{c}$)

\begin{equation}
A_\mu=sxW_{0\mu}^3+sW^{3}_{1\mu}+cB_{2\mu},
\end{equation}
\begin{equation}
Z_\mu =cxW_{0\mu}^3 +cW_{1\mu}^3-s B_{2\mu},
\end{equation}
\begin{equation}
Z_\mu^\prime=W_{0\mu}^3-xW_{1\mu}^{3}.
\label{eq:Z_prime}
\end{equation}
The charged gauge boson mass matrix can similarly be computed and given by
\begin{equation}
M_{C}^2=\frac{g_0^2F^2}{4}
\begin{pmatrix}
1&-x\\
-x&x^2(1+r^2)
\end{pmatrix}.
\end{equation}
The mass eigenvalues are
\begin{equation}
m_W^2=\frac{g_0^2F^2r^2x^2}{4}(1-x^2);\qquad m_{W^\prime}^2=\frac{g_0^2F^2}{4}(1+x^2).
\label{eq:chrg_mass}
\end{equation}
The mass eigenstates of the charged gauge bosons are
\begin{equation}
W_\mu=xW_{0\mu}+W_{1\mu},
\end{equation}
\begin{equation}
W_\mu^\prime=W_{0\mu}-xW_{1\mu},
\end{equation}
where the one with the smaller mass is identified as the $W$ boson and the heavier one as the $W^\prime$ boson. It is to be noticed that the massive SM gauge bosons get their contribution predominantly from $SU(2)_1$. However, for the BSM gauge bosons, $SU(2)_0$ is the major contributor. In the limit $g_0\rightarrow\infty$ (or equivalently $x\rightarrow0$), the $m_{W',Z'}$ decouple and the gauge spectrum at low energies thus reduces to that of the SM as expected. For all practical purposes $x$ can be parametrized as the ratio of the light and heavy gauge boson masses (to leading order) as
\begin{equation}
x^2\approx\frac{m_W^2}{r^2m^2_{W^\prime}}
\label{eq:xdef}
\end{equation}
which can be readily seen from Eqns.~\ref{eq:chrg_mass} and \ref{eq:neutral_boson_mass}.

After taking into account the gauge charges of the scalars and the imposed $Z_2$ symmetry, the most general scalar potential can be written as

\begin{align}
\begin{split}
V(\Phi_{1},\Phi_{2},\Sigma)&=m_2^2\,\textrm{Tr}(\Phi_2^\dagger\Phi_2)+\lambda_1\bigg[\Phi_1^\dagger\Phi_1-\frac{v^2}{2}\bigg]^2+\lambda_2\bigg[\textrm{Tr}(\Phi_2^\dagger\Phi_2)\bigg]^2\\
&+\lambda_{12}\bigg[\Phi_1^\dagger\Phi_1-\frac{v^2}{2}\bigg]\bigg[\textrm{Tr}(\Phi_2^\dagger\Phi_2)\bigg]+\lambda_{23}F^2\bigg[\textrm{Tr}(\Sigma\Phi_2\Sigma\Phi_2)\bigg]+\tilde\lambda_{23}F^2\bigg[\textrm{Tr}(\Sigma\Phi_2)\bigg]^2.
\end{split}
\end{align}
After spontaneous symmetry breaking the higgs mass ($m_h^2$) is given by the expression $2\lambda_1v^2$ and the charged scalar mass ($m_{\phi^\pm}^2$) is given by $m_2^2-\lambda_{23}F^2$. The two neutral components of $\Phi_2$ have masses given by the following expressions:
\begin{equation}
\label{equ:sca_mass}
m_{\phi_0}^2=m_2^2+\lambda_{23}F^2+2\tilde\lambda_{23}F^2,~~~~~m_{\phi_3}^2=m_2^2-\lambda_{23}F^2.
\end{equation}
It is important to note here that the $Z_2$ symmetry of the $\Phi_2$ bi-doublet works as the stabilising symmetry required for the DM under the WIMP scenario. However, this discrete symmetry can not forbid the number changing processes among the components of this bidoublet. Thus for a positive $(\lambda_{23}+\tilde\lambda_{23})$, the $\phi_3$ component will be the natural DM candidate and for negative values of this quantity, the $\phi_0$ component serves the same purpose from the stability argument. However, if the couplings are such that these two are degenerate, then we have an interesting possibility of a multicomponent DM scenario where both neutral components are cosmologically stable DM candidates - we describe how this will play out in more detail in Sec.~\ref{sec:dmpheno}.

The Yukawa sector is assumed to mimic that of the SM as $\Phi_1$ plays a similar role of the SM Higgs doublet. However, the fermions can still couple to the heavy gauge bosons due to gauge mixing. All the relevant couplings can be straightforwardly computed (again in a perturbative expansion in $x$) and are listed in Appendix~\ref{ap:coup}. Specifically, the purely scalar sector couplings are given in Tables \ref{tab:scalar_cubic} and \ref{tab:scalar_quartic}. The scalar couplings with the gauge bosons are listed in Tables \ref{tab:higgs_gauge_cubic},\ref{tab:higgs_gauge_quartic},\ref{tab:DM_gauge_quartic},\ref{tab:charged_scalar_DM_gauge_quartic},\ref{tab:charged_scalar_gauge_quartic},\ref{tab:charged_gauge_quartic} and \ref{tab:BSM_scalar_derivative_gauge_cubic}. The relevant couplings of the gauge bosons with the fermions can be found in Tables \ref{tab:BSM_gauge_quark_cubic} and \ref{tab:BSM_gauge_lepton_cubic} and that of the gauge bosons among themselves in Table \ref{tab:gauge}.


\section{Model Constraints}
\label{sec:constraints}
In this section, we impose all the theoretical and experimental constraints on the model. We will not only identify all the relevant parameters for carrying out further analysis, but also assign values to some and constrain the rest with various experimental observations. Among the many model parameters, the following are the most relevant for our analysis: $v$, $m_h$, $m_{\phi_3}$, $\lambda_{12}$, $\lambda_{23}$, $\tilde\lambda_{23}$, $m_{W^\prime}$ (or $m_{Z^\prime}$) and $F$ (or equivalently $r$).  The other important parameters can be represented in terms of these above mentioned ones. For example, $x$ can be represented in terms of $m_{Z^\prime}$ and $F$ following Eqn.~\ref{eq:neutral_boson_mass} or more conveniently through the mass ratio in Eqn.~\ref{eq:chrg_mass} (for a given $v$, $F$ and $r$ can be used interchangeably). We also set $m_h=125$ GeV and $v=246$ GeV in this paper. We now look at four key constraints on the model parameters coming from vacuum stability, the branching ratio of invisible decay of the Higgs, the Higgs alignment limit and collider constraints on the heavy gauge bosons.

\subsection{Vacuum Stability}
\label{subsec:vacstab}
It is the quartic part of the scalar potential that contributes to the vacuum stability conditions \cite{Bhattacharyya:2015nca}. After spontaneous symmetry breaking when $\Sigma$ develops a vacuum expectation value, the quartic part of the scalar potential in our model can be written in the unitary gauge $\langle\Sigma\rangle=1$  as,
\begin{equation}
V_4=\lambda_1(\Phi_1^\dagger\Phi_1)^2+\lambda_2[Tr(\Phi_2^\dagger\Phi_2)]^2+\lambda_{12}(\Phi_1^\dagger\Phi_1)[Tr(\Phi_2^\dagger\Phi_2)]
\end{equation}
The vacuum stability conditions can now be found through the standard procedure described in \cite{Bhattacharyya:2015nca}. Let's assume that $\Phi_1^\dagger\Phi_1=a$ \& $Tr(\Phi_2^\dagger\Phi_2)=b$. Then,
 
\begin{equation}
 V_4 =\lambda_1a^2+\lambda_2b^2+\lambda_{12}ab=(\sqrt{\lambda_1}a-\sqrt{\lambda_2}b)^2+(2\sqrt{\lambda_1\lambda_2}+\lambda_{12})ab.
\end{equation}

\begin{itemize}
\item Let us consider first the $b=0$ direction ($a\rightarrow\infty$). We demand $V_4\geq0$, which implies $\lambda_1a^2\geq0$. The condition is then $\lambda_1\geq0$.

\item Next we consider the $a=0$ direction ($b\rightarrow\infty$). Similarly by demanding $V_4$ to be positive, one can derive the condition $\lambda_2\geq0$.

\item Lastly, the direction $a=\sqrt{\frac{\lambda_2}{\lambda_1}}b$ (where $a,b\rightarrow\infty$) can be considered. Again demanding $V_4\geq0$ one can find the condition $\lambda_{12}\geq-2\sqrt{\lambda_1\lambda_2}$.

\end{itemize}
Now collecting all the equations together the vacuum stability conditions for this model are given by
\begin{equation}
\lambda_1\geq0,\qquad \lambda_2\geq0 \qquad, \textrm{and}\qquad \lambda_{12}\geq-2\sqrt{\lambda_1\lambda_2}.
\end{equation}
Another relevant theoretical constraint, the perturbativity bound, is discussed in Appendix \ref{ap:per}.

\subsection{Higgs Invisible Decay}
\label{subsec:higgsinvdec}
\begin{figure}[h!]
\centering
\includegraphics[scale=0.7]{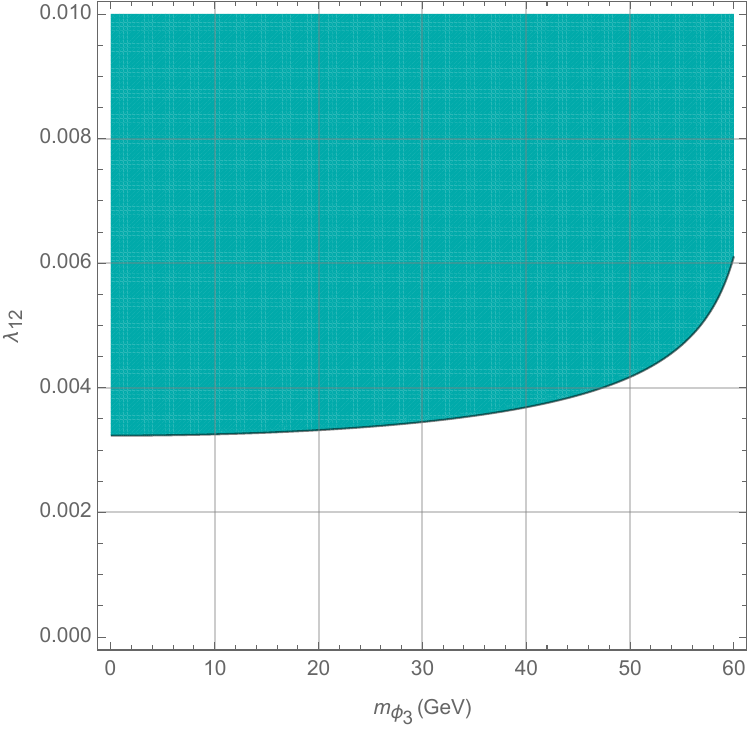}
\caption{The shaded region in the $\lambda_{12}-m_{\phi_3}$ space is disallowed by constraints from Higgs invisible branching ratio.}
\label{fig:higgsinvdec}
\end{figure}

The Higgs invisible decay branching ratio is given by
\begin{equation}
 BR_{h\rightarrow inv}=\frac{\Gamma_{h\rightarrow inv}}{\Gamma_{h\rightarrow inv}+\Gamma_{h\rightarrow SM}}.
\end{equation}
Among ATLAS \cite{ATLAS-CONF-2020-052} and CMS \cite{CMS:2018yfx} measurements of this quantity, the more stringent bound comes from the ATLAS experiment  and is  11\%. In Fig.~\ref{fig:higgsinvdec}, assuming $\phi_3$ to be the dark matter candidate, we present this bound in the parameter space $m_{\phi_3}-\lambda_{12}$ for dark matter masses smaller than half of $m_h$ (in order, of course, that $h$ can decay into a pair of them). There is an upper bound on $\lambda_{12}$ as a function of $m_{\phi_3}$ as indicated by the shaded region in the figure. For higher DM masses the restriction will be drastically lenient due to off-shell suppression of the decay width.

\subsection{Higgs Alignment Limit}
\label{subsec:align}

\begin{figure}[h!]
\centering
\includegraphics[scale=0.4]{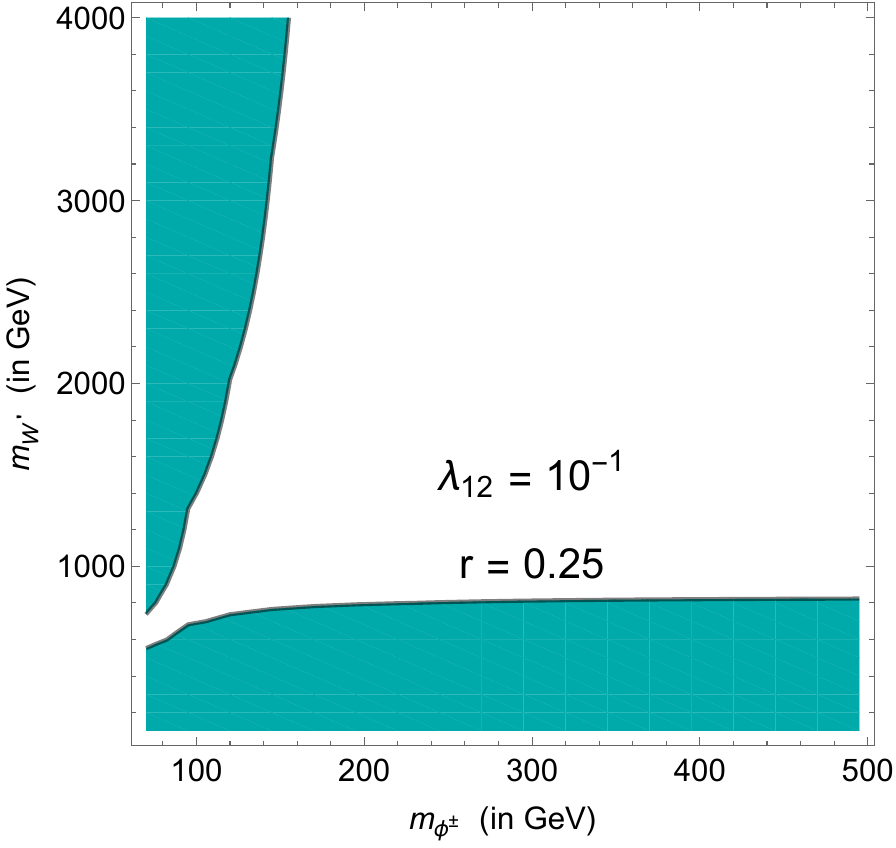}
\hspace{0.1cm}
\includegraphics[scale=0.38]{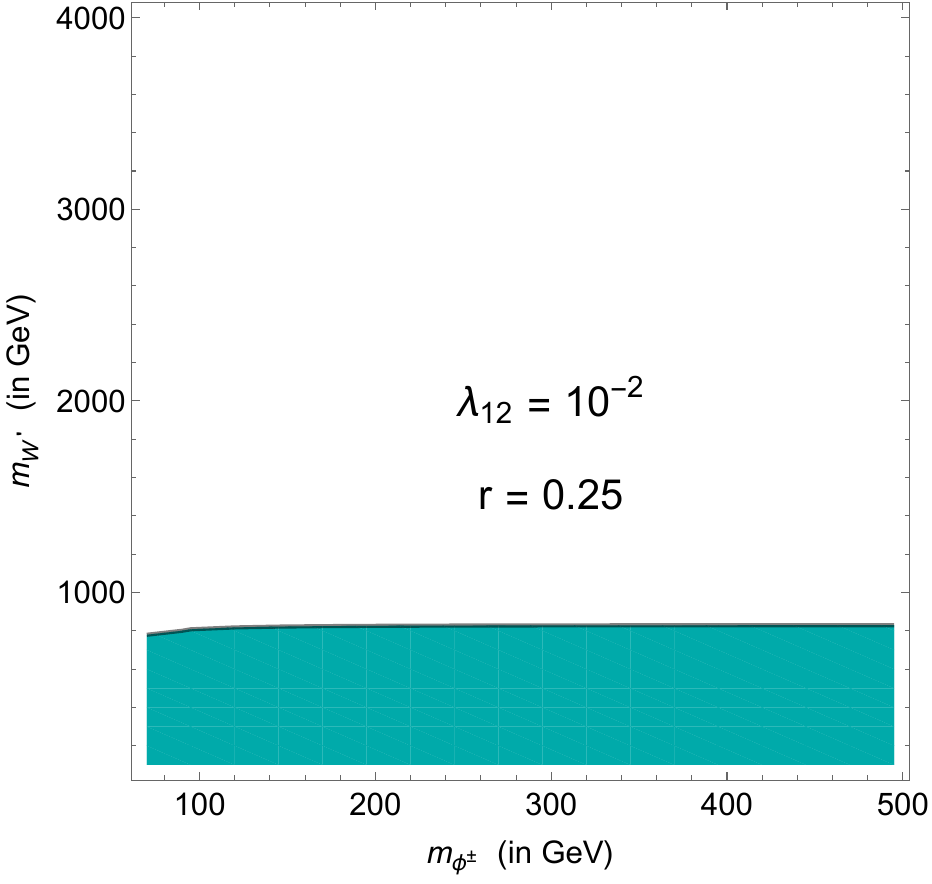}
\caption{The dark cyan shaded region in the $m_{\phi^{\pm}}$ vs.~$m_{W^{'}}$ plane is disallowed from the branching ratio measurements of $h \to \gamma \gamma$ channel.}
\label{Fig:h2diphoton}
\end{figure}

Given the extended gauge and scalar sector in our model, the standard Higgs boson couplings to various constituents of the SM are modified due to gauge mixing. For example, the coupling between Higgs and the pair of $W$ boson (Table \ref{tab:higgs_gauge_cubic} ) will take the form 
\begin{equation}
\lambda_{hW^{+}_{\mu}W^{-}_{\mu}} = \frac{e^{2}v}{2\sin^{2}\theta_{w}}\left(1 + x^{2}\sin^{2}\theta_{w}\right) + h.c. 
\label{eq:hWW}
\end{equation} 
One can notice that the deviation of this coupling from the SM counterpart is proportional to $x^{2}$. A similar deviation can be found in other SM-like gauge couplings in Tables \ref{tab:higgs_gauge_cubic} and \ref{tab:higgs_gauge_quartic}. On the other hand the couplings between the SM Higgs and different fermion pairs in this model maintain the same structure as SM due to the Yukawa sector of both these models being the same. To analyse how new physics can modify Higgs properties we look into the branching ratio of $h \to \gamma \gamma$. Apart from the usual SM fermion and $W$ boson loops, the heavy $W^{'}$ boson as well as the charged scalar $\phi^{\pm}$ will contribute to this loop mediated process. After incorporating these additional contributions we have calculated the branching ratio $\text{BR}\left(h \to \gamma \gamma\right)$ (for a detailed calculation, see Appendix~\ref{ap:h_gamma_gamma}) in this model for a range of charged Higgs and $W^\prime$ masses. In Fig.~\ref{Fig:h2diphoton}, we present the allowed region (shaded region is disallowed) of the parameter space after recasting ATLAS \cite{ATLAS:2022tnm} and CMS \cite{CMS-PAS-HIG-19-015} measurements - we have set the ratio between two \emph{vev's} $r$ at 0.25 and presented our results for two different $\lambda_{12}$ benchmark values given by 0.1 and 0.01. It is found that for both the scenarios, the region $m_{W^\prime}>$ 1 TeV and $m_{\phi^{\pm}}>$ 200 GeV is allowed.    

\subsection{Collider Constraints on Heavy Gauge Bosons}
\label{subsec:colcon}
Due to its extended gauge sector, the model admits heavy charged and neutral gauge bosons. The $Z^\prime$ couples to different SM fields primarily through gauge mixing. As a result, this $Z^\prime$, if produced in colliders, can be discovered from its decays to different SM final states. Specifically, the $Z^\prime$ can decay into $\ell_{L}\bar{\ell}_{L}$ ($\ell$ represents all possible charged leptons), $\nu_{L}\bar{\nu}_{L}$ and $q_{L}\bar{q}_{L}$ ($q$ represents all possible SM coloured particles) pairs where the the corresponding coupling is suppressed by the parameter $x$ \emph{w.r.t} the SM counterpart. One can notice in Tables \ref{tab:BSM_gauge_quark_cubic} and \ref{tab:BSM_gauge_lepton_cubic} that the $Z^\prime$ specifically couples to the left-handed chiral fermions since it is induced purely from a gauge mixing in the neutral gauge sector and does not receive any contribution from the $U(1)_2$ (see Eqn.~\ref{eq:Z_prime}). In addition to the fermion pairs, the $Z^\prime$ can decay into two other SM final states $Zh$ and $WW$. Apart from these SM final states the $Z^\prime$ boson does couple to the dark sector particles $\phi^{+}\phi^{-}$ and $\phi_{0}\phi_{3}$. The three point vertices corresponding to these two decay modes are proportional to $\frac{1}{x}$ at leading order (see Table \ref{tab:BSM_scalar_derivative_gauge_cubic}). As a consequence the decay width of these two modes are significantly larger than SM final states. In Table \ref{tab:Zpdecay} we present the analytic formulas for these decay widths.

\begin{figure}[h!]
\centering
\includegraphics[scale=0.25]{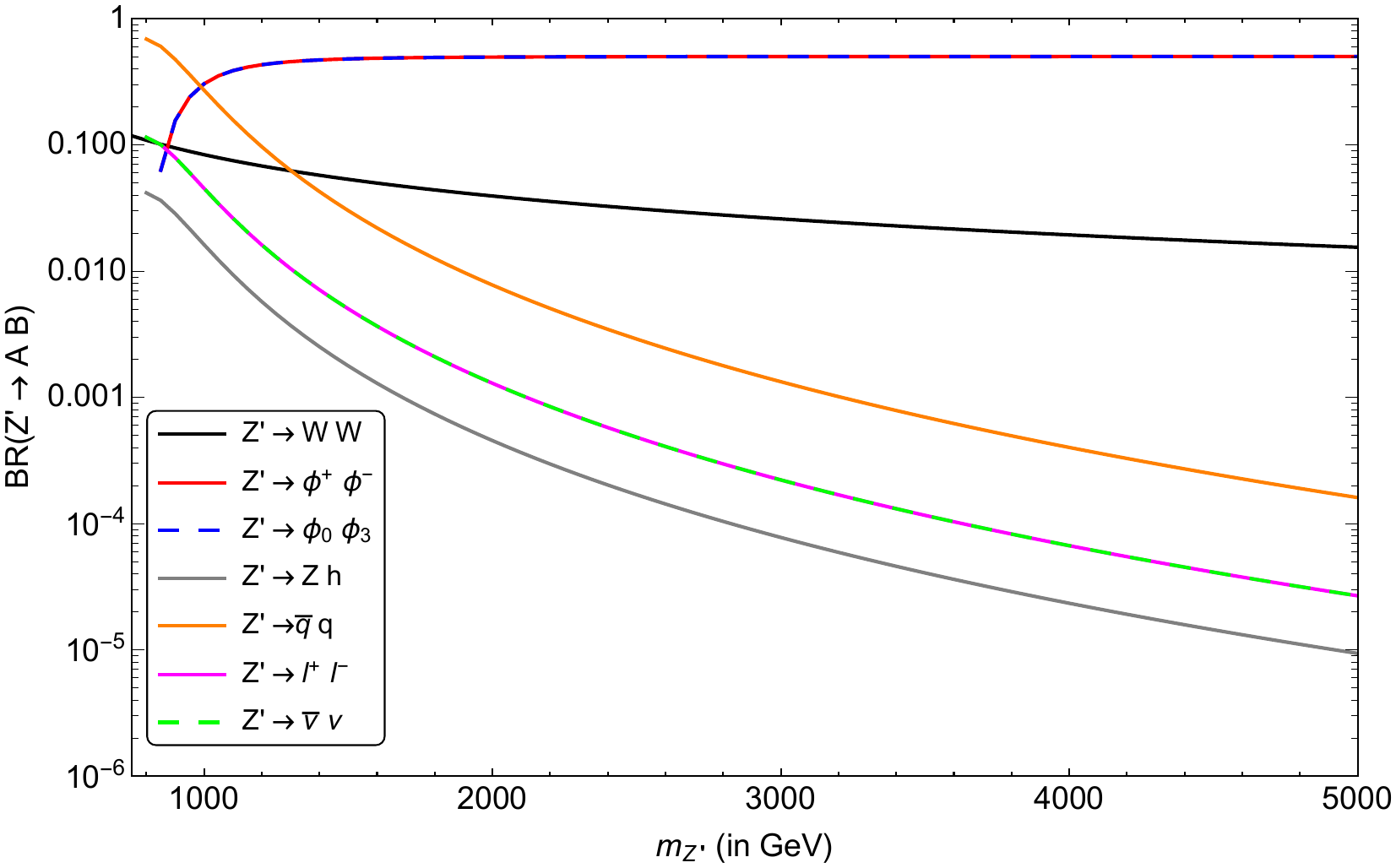}
\includegraphics[scale=0.37]{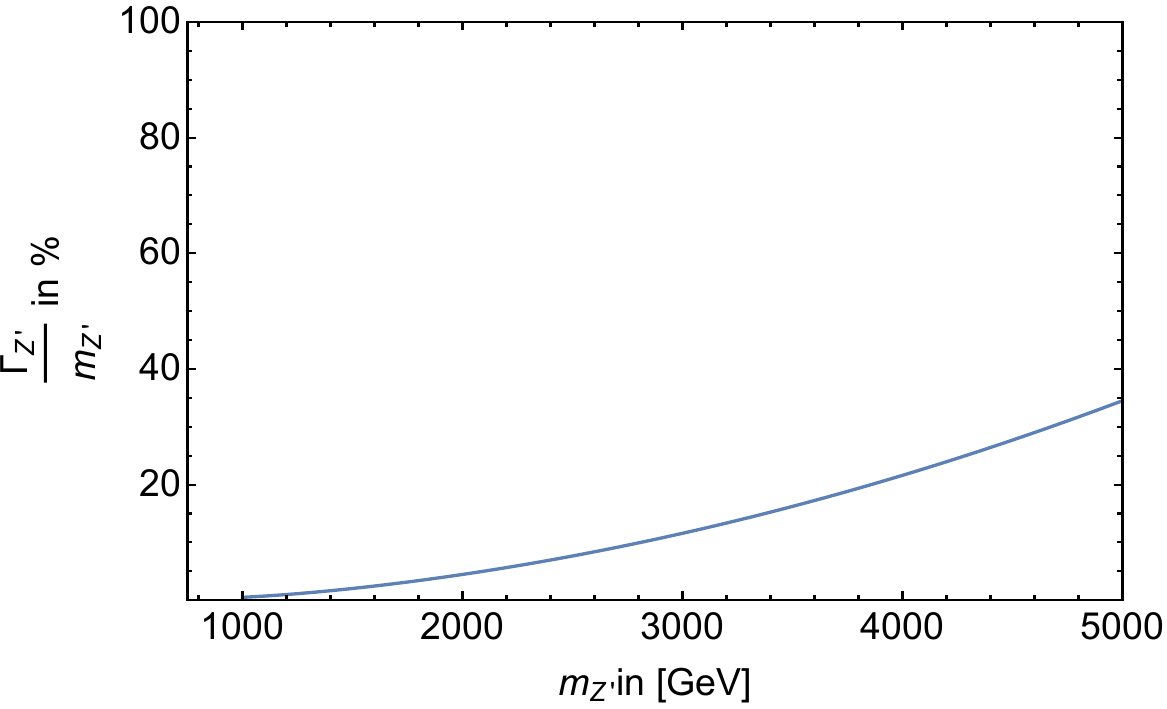}
\caption{Left: The branching ratio of $Z^{'}$ for all possible two body final states. Here we have set $r$ = 0.25 and $m_{\phi_{0}}=m_{\phi_{3}} = m_{\phi^{\pm}}$ = 400 GeV. The branching ratio for $Z^{'} \to q q~$ mode (solid orange line) is summed over all six SM quark final states. Similarly, the $Z^{'} \to \ell^{+} \ell^{-}$ (blue solid) and $Z^{'} \to \nu \nu~$ (green dashed) modes are calculated while considering all three lepton flavours. Right: The ratio $\frac{\Gamma_{Z^{'}}}{m_{Z^{'}}}$  for $m_{Z^{'}}$ mass ranging from 1 TeV to 5 TeV. Here $\Gamma_{Z^{'}}$ is the total decay width which is computed considering all possible two body decay modes.}
\label{fig:Zppheno}
\end{figure}

\begin{table}[h!]
\begin{center}
\begin{tabular}{|c|c|}
\hline
Decay Modes& Decay rate expressions \\
\hline
\rule{0pt}{4ex}
$\Gamma\left(Z^{'} \to f\bar{f}\right)$&$ \frac{N_{c}\lambda^2_{Z^{'}f\overline{f}}}{48\pi}m_{Z^{'}} $\\
\hline
\rule{0pt}{4ex}
$\Gamma\left(Z^{'} \to W^{+}W^{-}\right)$ &$ \frac{\lambda^{2}_{WWZ^{'}}m_{Z^{'}}}{192\pi}\sqrt{1 - \frac{m^{2}_{W}}{m^{2}_{Z^{'}}}}\left(1 + 16\frac{m^{2}_{W}}{m^{2}_{Z^{'}}} - 68\frac{m^{4}_{W}}{m^{4}_{Z^{'}}} - 48\frac{m^{6}_{W}}{m^{6}_{Z^{'}}}\right)$ \\
\hline
\rule{0pt}{5ex}
$\Gamma\left(Z^{'} \to Zh\right)$ &$ \frac{\lambda^{2}_{hZZ^{'}}}{48\pi m_{Z^{'}}}\left(2 + \frac{\left(m^{2}_{Z^{'}} + m^{2}_{Z} - m^{2}_{h}\right)^{2}}{4m^{2}_{Z}m^{2}_{Z^{'}}}\right)\sqrt{1 + \frac{\left(m^{2}_{h} - m^{2}_{Z}\right)^{2}}{m^{4}_{Z^{'}}} - 2\frac{\left(m^{2}_{h} + m^{2}_{Z}\right)^{2}}{m^{4}_{Z^{'}}}}$\\
\hline
\rule{0pt}{5ex}
$\Gamma\left(Z^{'} \to \phi_{0}\phi_{3}\right)$ &$ \frac{\lambda^{2}_{\phi_{0}\phi_{3}Z^{'}}}{48\pi m_{Z^{'}}}\left(m^{2}_{Z^{'}} - 2\left(m^{2}_{\phi_{0}} + m^{2}_{\phi_{3}}\right)\right)\sqrt{1 + \frac{\left(m^{2}_{\phi_{0}} - m^{2}_{\phi_{3}}\right)}{m^{4}_{Z^{'}}} - 2\frac{\left(m^{2}_{\phi_{0}} + m^{2}_{\phi_{3}}\right)}{m^{2}_{Z^{'}}}}$ \\
\hline
\rule{0pt}{5ex}
$\Gamma\left(Z^{'} \to \phi^{+}\phi^{-}\right)$ & $ \frac{\lambda^{2}_{\phi^{+}\phi^{-}Z^{'}}}{192\pi m_{Z^{'}}}\left(m^{2}_{Z^{'}} - 4m^{2}_{\phi^{\pm}}\right)\sqrt{1 - 4\frac{m^{2}_{\phi^{\pm}}}{m^{2}_{Z^{'}}}}$\\
\hline
\end{tabular}
\caption{The relevant $Z^\prime$ decay rates for the purposes of this section - the expressions for the couplings given herein can be found in Appendix~\ref{ap:coup}.}
\label{tab:Zpdecay}
\end{center}
\end{table}
Here $N_{c}$ stands for the colour factor (3 and 1 for quarks and leptons respectively) and the $\lambda$'s denote the relevant cubic vertices. We have neglected the mass of the fermions in comparison with the $Z^\prime$ mass for the fermionic decay modes. In principle, there exists a cubic vertex of the form $\lambda_{Z^{'}WW^{'}}$. However, in the present model $m_{Z^{'}}$ and $m_{W^{'}}$ are degenerate and the $Z^{'}$ decay via $WW^{'}$ channel is kinematically forbidden. In Fig.~\ref{fig:Zppheno} (left) we present the corresponding BRs of these channels for $m_{Z^{'}}$ranging from 1 TeV to 5 TeV. As expected, the BR of the $\phi^{+}\phi^{-}$ and $\phi_{0}\phi_{3}$ modes dominate for the entire $m_{Z^{'}}$ range. Among the SM final states the $WW$ has the maximum and $Zh$ the minimum branching ratio. For the present calculation we have set $r$ = 0.25 and $m_{\phi_{0}}=m_{\phi_{3}} = m_{\phi^{\pm}}$ = 400 GeV. In Fig.~\ref{fig:Zppheno} (right) we present the $\frac{\Gamma_{Z^{'}}}{m_{Z^{'}}}$ in percentage. There is a visible growth in this ratio at the mass range $m_{Z^{'}}>$ 2.5 TeV. However, the overall value remains below $\lesssim$ 30$\%$ for the entire range. This is to emphasize that for the current benchmark choice the perturbative treatment is applicable. For different parameter choices, one should explicitly calculate the $\frac{\Gamma_{Z^{'}}}{m_{Z^{'}}}$ and ensure that the corresponding value is less than unity.   

\begin{figure}[h!]
\centering
\includegraphics[scale=0.3]{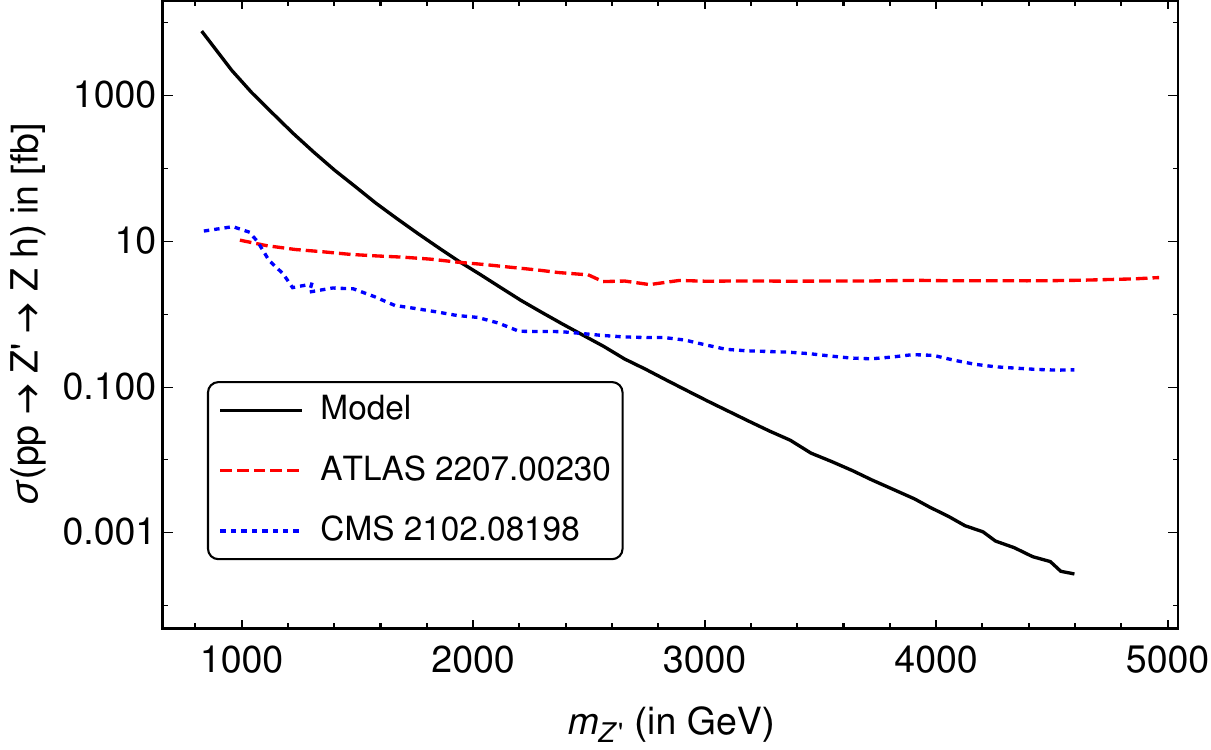}
\includegraphics[scale=0.3]{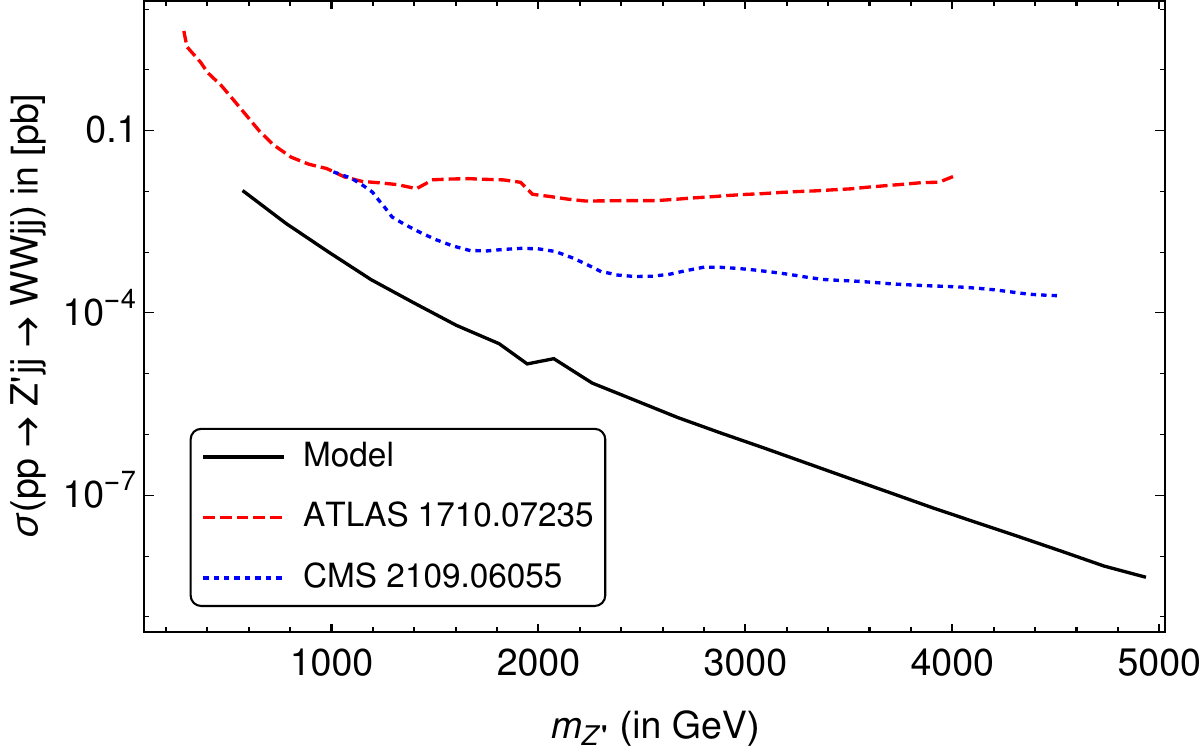}
\includegraphics[scale=0.3]{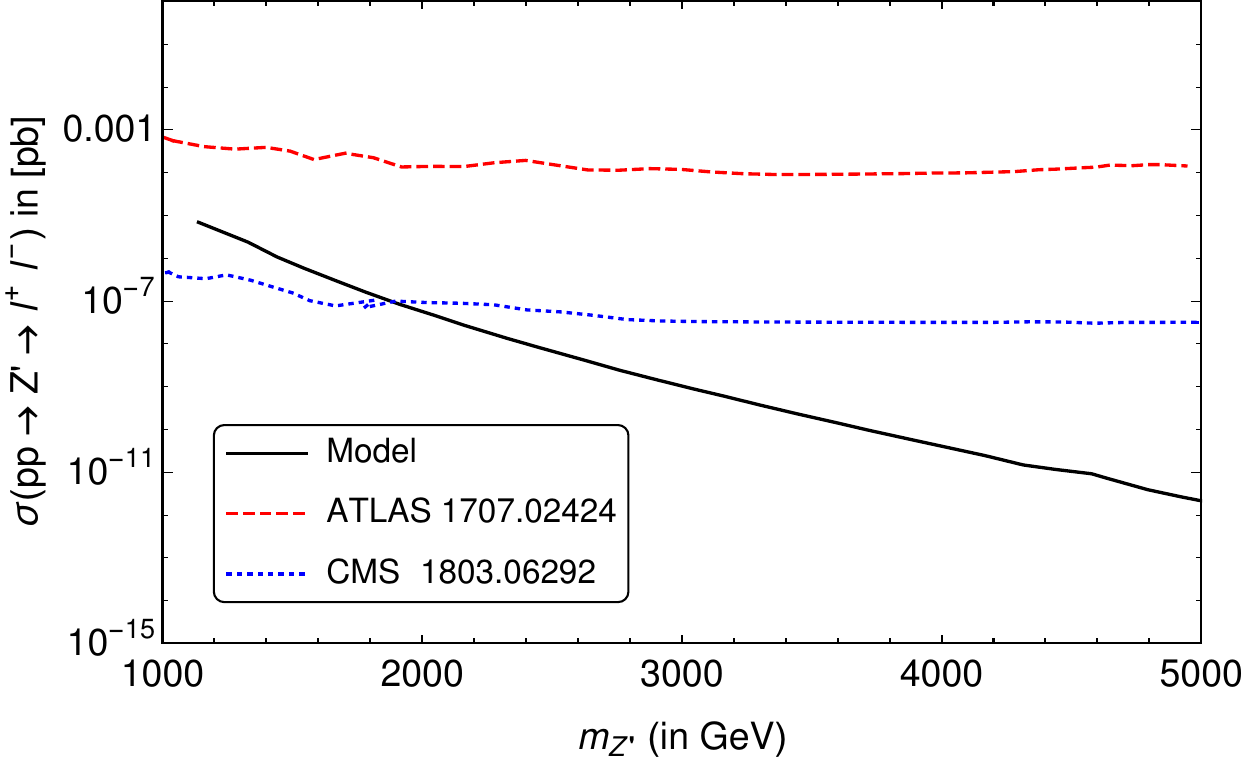}
\caption{The direct search limits in the channels $\text{DY} \to Z^{'} \to Zh$ (upper panel left), $\text{VBF} \to Z^{'}jj \to WWjj$ (upper panel right) and $\text{DY} \to \ell^{+} \ell^{-}$ (lower panel). The red-dashed, blue-dotted and black solid line denote the ATLAS, CMS and the model specific curves respectively. The arXiv IDs which are written beside the ATLAS and CMS highlight the corresponding articles \cite{CMS:2021fyk,ATLAS:2022enb,ATLAS:2017jag,CMS:2021klu,ATLAS:2017fih,CMS:2018ipm} which are used in this calculation.}
\label{fig:Zpbound}
\end{figure}

From the above discussion, it is evident that the $Z^{'}$ in this model primarily decays via exotic scalars coming from the scalar bidoublet and that the branching ratios in the SM final states are significantly low. Nevertheless, it is useful to check the direct detection limits to understand the allowed parameter space. Before the LHC era, the LEP and Tevatron colliders have extensively searched for the heavy neutral gauge bosons. The LEP searches \cite{Electroweak:2003ram} can be divided into two categories: i) precision measurement around the $Z$ pole which can effectively constraint the mixing angle between the $Z - Z^{'}$ and ii) direct searches for the heavy neutral boson in the $e^{+} e^{-} \to f \bar{f}$ (here $f$ is all possible SM fermions) at the mass range above $Z$-pole. On other hand, at Tevatron the $Z^{'}$ boson was primarily looked for in the $p\bar{p} \to Z^{'} X \to \ell^{+} \ell^{-} X$ channel and using the data, both the CDF\cite{CDF:1994kud,CDF:1997wdd} and D0 \cite{D0:1996qsx,D0:2001zjx} collaborations have put bounds on $m_{Z^{'}}$ for different $U(1)$ extensions of the SM. The commonality between the LEP and Tevatron is the corresponding bounds are strictly dependent on the $Z^{'}$ couplings to SM fermions. For a comprehensive analysis on how these results can be translated in a specific model, the interested reader can consult Ref.~\cite{Carena:2004xs,Bandyopadhyay:2018cwu}. At LHC, both the Drell-Yan (DY) and Vector Boson Fusion (VBF) production mechanisms are considered for the $Z^{'}$ searches. After production, the $Z^{'}$ can decay to dileptons, $Zh$, and $WW$ final states. For the present study, we will consider these three final states and will recast the corresponding ATLAS and CMS bounds in our model. To calculate the signal cross section in our model we have used the prescription illustrated in Ref.~\cite{Pappadopulo:2014qza,Florez:2016uob}. We begin with the process $pp \to Z^{'} Zh$, where the $Z^{'}$ is produced via DY and decays into $Zh$. The $Z$ further decays leptonically and the Higgs boson decays to $\bar{b}b$. Both the ATLAS \cite{ATLAS:2022enb} and CMS \cite{CMS:2021fyk} have searched the $Z^{'}$ boson in this mode. The model dependent upper bounds on the $\sigma\times\text{BR}$ is displayed in Fig.~\ref{fig:Zpbound} (upper panel left) by red dashed (ATLAS) and blue dotted curve (CMS) curves. The black solid line represents the corresponding number for the present model.  

In the case of VBF mechanism the $Z^{'}$ boson can decay via $WW$ mode. After that, one of the $W$ bosons decays hadronically and the other one decays leptonically. In Fig.~\ref{fig:Zpbound} (upper panel right), we display the corresponding ATLAS, CMS and model specific $\sigma\times\text{BR}$ values via red dashed, blue dashed and black solid line respectively. We have also translated the LHC limits in dilepton final state where the $Z^{'}$ is produced via DY mechanism. The corresponding limits \cite{ATLAS:2017fih,CMS:2018ipm} are displayed in Fig.~\ref{fig:Zpbound} (lower panel). Considering these search channels one can see that the most stringent bound can be obtained from the $Zh$ mode and to satisfy current collider limits one have to set $m_{Z^{'}} \gtrsim$ 2.5 TeV for the specific benchmark points that we have chosen. Apart from the heavy neutral gauge bosons the LHC also looked for the additional charged gauge bosons. However the number of searches is comparatively lower as well as less stringent in the latter case.


\section{Dark Matter Phenomenology}
\label{sec:dmpheno}
In this section we consider our model described in Sec.~\ref{sec:model} as a plausible solution to the dark matter problem. In what follows, we discuss the choice of the DM candidate, calculate the relic abundance for a multicomponent scenario, and address the direct detection constraints.

\subsection{Choice of Dark Matter Candidate} 
\label{subsec:dm_can}

For generic choices of parameters in this model, the couplings mentioned in Table \ref{tab:BSM_scalar_derivative_gauge_cubic} allow the  decays $\phi_0\rightarrow\phi^\pm W^{\prime\mp},\phi_3Z^\prime$ for a positive $\lambda_{23}+\tilde\lambda_{23}$, and $\phi_3\rightarrow\phi_0Z^\prime$ for negative values of $\lambda_{23}+\tilde\lambda_{23}$. Such decays, despite being suppressed by the heavy gauge boson mass, would still make one of the scalars unstable. In other words, constraining such decays with the age of universe would push the heavy gauge boson masses to very high values essentially decoupling them from SM gauge sector for all practical purposes. While this is certainly a reasonable path to take, in this work our aim is to treat the multi-component DM scenario while maintaining a meaningful way of discussing the gauge sector in the context of present experiments and constraints. Thus, we work under the mass degeneracy limit of the $Z_2$ odd sector that makes the neutral components cosmologically stable DM candidates. Referring to Eqn.~\ref{equ:sca_mass}, we see that this corresponds to the particular parameter choice $\tilde\lambda_{23}=-\lambda_{23}$. The charged components will completely annihilate via photon couplings. For the mass degenerate scenario, all components of the $Z_2$ odd sector are stable at tree level in the context of collider production as well. However, in the absence of higher dimensional terms in our scalar Lagrangian, loop level effects can give rise to a sub-GeV mass non-degeneracy between the charged and neutral components \cite{Cirelli:2005uq} leading to charged scalar decays via MeV scale charged pions as in reference \cite{Belyaev:2016lok}, thus avoiding any constraints pertaining to the production of an otherwise stable charged particle in colliders which could potentially disallow charged scalar masses (and thus the DM masses) upto a few TeV. Given that the present model admits an extra $SU(2)$ gauge group in addition to the SM one, we expect such loop effects to be even more pronounced in our case in providing a mass splitting between the charged and the neutral scalars.

\subsection{Relic Abundance}
\label{subsec:relic}
In the early universe all the particles were in equilibrium in the high temperature cosmic soup. Over time as the universe expanded and the temperature of the universe started to drop, the interaction rate of WIMPs with the other particles of the cosmic plasma became comparable to the Hubble parameter. It became difficult for the WIMPs to find another WIMP to annihilate with and thus they were frozen out remaining in the universe as a relic which one can compare with the observed DM relic abundance. The value of relic abundance as measured by Planck satellite is given by $\Omega h^2=0.120\pm0.001$ \cite{planck2018}.

\begin{figure}[h!]
\centering
\includegraphics[scale=0.6]{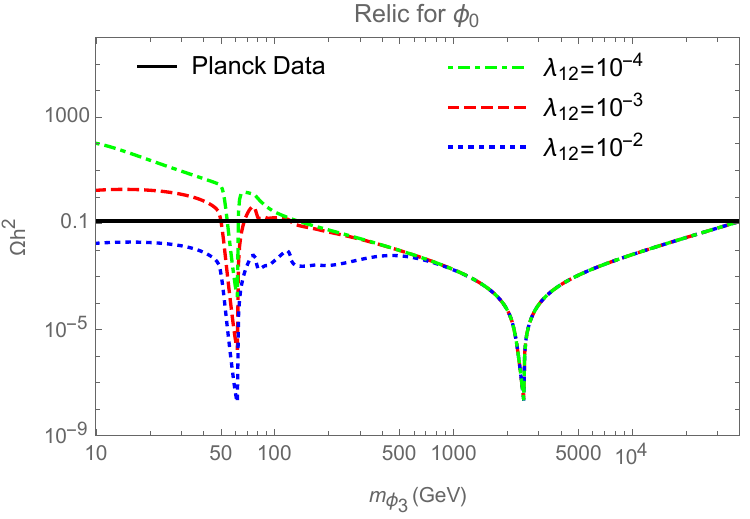}\includegraphics[scale=0.6]{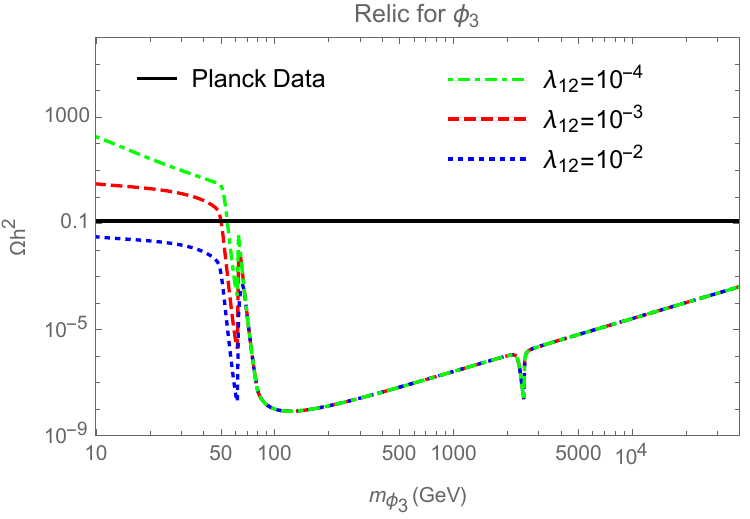}\\
\includegraphics[scale=0.6]{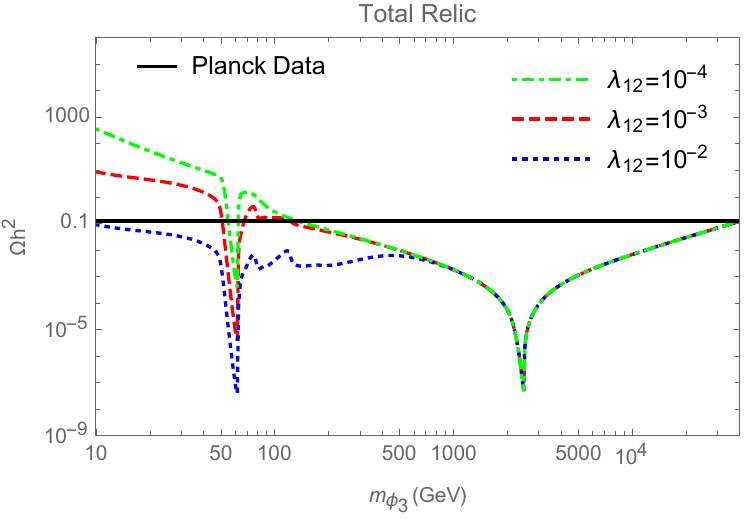}
\caption{Relic abundance as a function of dark matter mass for various couplings (lagrangian parameters) in the multi-component dark matter scenario with annihilation via $h$ portal and co-annihilation mostly via $Z'$ portal for $m_{Z^\prime}=5$ TeV.}
\label{fig:relic_final}
\end{figure}

There are two DM candidates in this model as argued in Sec.~\ref{subsec:dm_can}. The dark matter candidates can annihilate mainly via higgs mediated channels to SM states, for example $b\bar b$, $\tau\bar\tau$, $W^+W^-$, $ZZ$, $hh$ and $t\bar t$, as allowed by the kinematics in various DM mass range. The Feynman diagrams are given in Appendix \ref{ap:beq}. Among the individual contributions of the DM candidates to the relic, the contribution from $\phi_3$ is much less than that of $\phi_0$ for the DM mass range starting from $m_W$ upto around 300 GeV. This is due to the fact that in this model there exists a $\phi_3\phi_3W^+W^-$ coupling (see Table \ref{tab:DM_gauge_quartic}) and also $\phi_3\phi^\pm W^\mp$ couplings (see Table \ref{tab:BSM_scalar_derivative_gauge_cubic}) which is absent for $\phi_0$, and thus $\phi_3$ annihilates more through those channels leaving less relic in the universe. These couplings come into effect around a DM mass of $m_W$ and finally go away as the co-annihilation channels start to dominate at a DM mass of around 300 GeV.\footnote{The exact DM mass value can be different than 300 GeV depending upon relevant couplings and the heavy gauge boson mass.}  As we move into the higher DM mass range, co-annihilation among $\phi_0$ and $\phi_3$ via $Z^\prime$ mediated diagrams into $WW$, $Zh$ and quark and lepton pairs start to dictate the relic abundance. All these properties are evident in Fig.~\ref{fig:relic_final} where the relic abundance contribution of each dark matter candidate as well as the total contribution have been displayed as a function of dark matter mass for a mass degenerate DM scenario. The $Z^\prime$ mass is fixed at 5 TeV as a benchmark choice and the relic varies with scalar sector parameter $\lambda_{12}$ in the annihilation dominated low DM mass region whereas at high DM mass region the $\lambda_{12}$ dependence is absent since the $Z^\prime$ mediated co-annihilation starts dictating the relic. There are two peaks in the velocity averaged cross section due to the Breit-Wigner resonance which are related to the mass of the two mediators $h$ and $Z^\prime$. DM relic is dictated by various annihilation and co-annihilation channels in various DM mass ranges:
\begin{itemize}
\item The DM annihilation to $\bar{b}b$ is the dominant channel dictating the relic in the DM mass range upto around the $W$ boson mass i.e.~$m_W\sim80.4$ GeV. The $\bar{\tau}\tau$ is the second most dominant process in this DM mass region. The amplitude for the process to $\bar{b}b$ final state is given by
\begin{equation}
|\mathcal{M}|_{\bar{b}b}^2\sim\frac{\lambda_{h\phi_i\phi_i}^2\lambda_{h\bar{b}b}^2}{(s-m_h^2)^2+m_h^2\Gamma_h^2}.
\end{equation}
\item One can observe the sudden decline in the relic curve for both the individual contribution and the total relic around the DM mass of $m_W$ as the Higgs mediated annihilation channel to $WW$ final state becomes kinematically accessible. The amplitude for the same is given by
\begin{equation}
|\mathcal{M}|_{WW}^2\sim\frac{\lambda_{h\phi_0\phi_0}^2\lambda_{hWW}^2}{(s-m_h^2)^2+m_h^2\Gamma_h^2}\bigg(1-\frac{4m_W^2}{s}+\frac{12m_W^4}{s^2}\bigg)
\end{equation}
for $\phi_0$ and
\begin{align}
\begin{split}
&|\mathcal{M}|_{WW}^2\sim\\
&\bigg(\frac{\lambda_{h\phi_3\phi_3}^2\lambda_{hWW}^2}{(s-m_h^2)^2+m_h^2\Gamma_h^2}+\lambda_{\phi_3\phi_3WW}^2+\frac{2\lambda_{h\phi_3\phi_3}\lambda_{hWW}\lambda_{\phi_3\phi_3WW}(s-m_h^2)}{(s-m_h^2)^2+m_h^2\Gamma_h^2}\bigg)\bigg(1-\frac{4m_W^2}{s}+\frac{12m_W^4}{s^2}\bigg)\\
&+\frac{4\lambda_{\partial(\phi^+)\phi_3W^-}^4}{s^2\big(\sqrt{s-4m_{\phi_3}^2}\sqrt{s-4m_W^2}\cos{\psi}+2m_W^2-s\big)^2}\times\\
&\bigg(s\cos{\psi}^2(s-4(m_{\phi_3}^2+m_W^2))-2s\sqrt{s-4m_{\phi_3}^2}\sqrt{s-4m_W^2}\cos{\psi}-16m_{\phi_3}^2m_W^2\sin{\psi}^2+s^2\bigg)^2\\
&+\textrm{the interference term}
\end{split}
\end{align}
for $\phi_3$.
\item There is a further sudden decline in the relic is evident in the $\phi_0$ contribution and the total relic around the Higgs mass (i.e.~$m_h\sim125$ GeV) as the annihilation to $hh$ final state becomes dominant. The case for $\phi_3$ contribution to relic is slightly different as mentioned above. The amplitude goes as
\begin{equation}
|\mathcal{M}|_{hh}^2\sim\bigg(\lambda_{hh\phi_i\phi_i}+\lambda_{h\phi_i\phi_i}^2\big(\frac{1}{u-m_{\phi_i}^2}+\frac{1}{t-m_{\phi_i}^2}\big)+\frac{\lambda_{h\phi_i\phi_i}\lambda_{hhh}}{s-m_h^2}\bigg)^2.
\end{equation}
\item Finally, at DM masses higher than a few hundred GeV the $Z^\prime$ mediated co-annihilation cross-section takes over and the $Z^\prime$ mediated channel to $WW$ final state emerges as the dominant contribution. The corresponding amplitude is proportional to
\begin{align}
\begin{split}
|\mathcal{M}|_{WW}^2\sim&\frac{\lambda_{WWZ^\prime}^2\lambda_{(\partial\phi_0)\phi_3Z^\prime}^2}{(s-m_{Z^\prime}^2)^2+m_{Z^\prime}^2\Gamma_{Z^\prime}^2}\frac{(s-4m_W^2)(s-4m_{\phi_i}^2)}{8m_W^4}\\
&\bigg(s^2+12sm_W^2+12m_W^4+\cos(2\psi)\big(s^2-4sm_W^2+12m_W^4\big)\bigg),
\end{split}
\end{align}
\end{itemize}
where $\psi$ is the angle between the incident and emergent particle momenta to be summed over. The region above the black line is disallowed by overabundance as per Planck data. The relevant Boltzmann equations are given in Appendix \ref{ap:beq}.

\begin{figure}[h!]
\begin{center}
\includegraphics[scale=0.65]{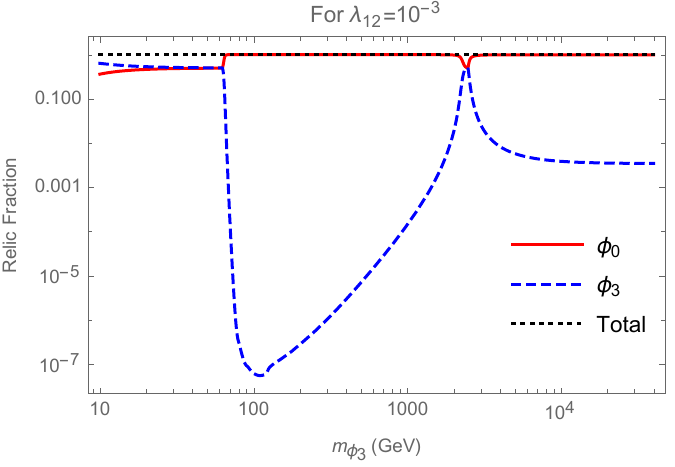}\includegraphics[scale=0.65]{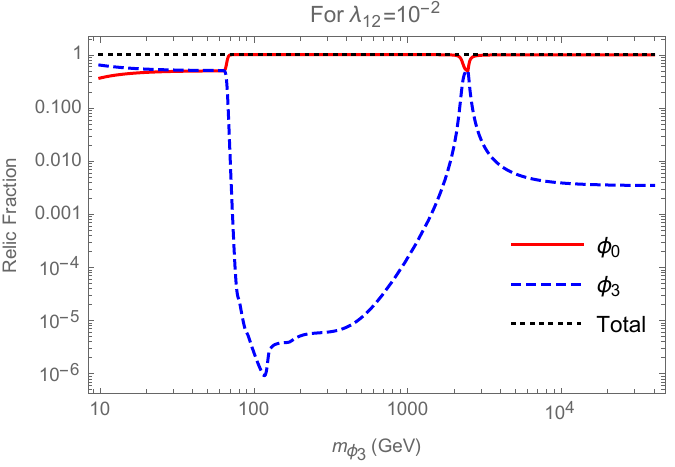}
\caption{Contribution of each component to the relic abundance with benchmark values same as in Fig.~\ref{fig:relic_final} and $\lambda_{12}=10^{-3}$ (Left), $10^{-2}$ (Right).}
\label{fig:relic_frac}
\end{center}
\end{figure}

The relative contribution of the individual DM components to the total relic abundance can be defined as a relic fraction given by $\frac{\Omega_{\phi_i}}{\Omega_{\phi_0}+\Omega_{\phi_3}}$ for the $\phi_i$ component. The relic fraction of each of the components is shown in Fig.~\ref{fig:relic_frac}. One can observe that the contribution to the relic from individual components are of the same order for DM masses $<$ 80 GeV due to similar diagrams contributing to both of them. For DM masses $>$ 80 GeV, the diagrams with $WW$ as final states become kinematically accessible. The diagrams shown in Fig.~\ref{fig:phi_3_anni} only exist for the $\phi_3$ component and are absent for the $\phi_0$ component. Hence, the $\phi_3$ component annihilates dominantly through these channels, especially through the $\phi_3\phi_3WW$ vertex, leading to a very small contribution to the relic. The $\phi_0$ on the other hand furnishes the major contribution to the relic. As one explores higher DM masses, the $Z^\prime$ mediated diagrams start to become relevant. The contributions are again comparable at the Breit-Wigner resonance region corresponding to the $Z^\prime$ mediator (i.e. around $m_{\phi_3}\sim\frac{m_{Z^\prime}}{2}\sim2.5$ TeV).

\begin{figure}[h!]
\centering
\includegraphics[scale=0.6]{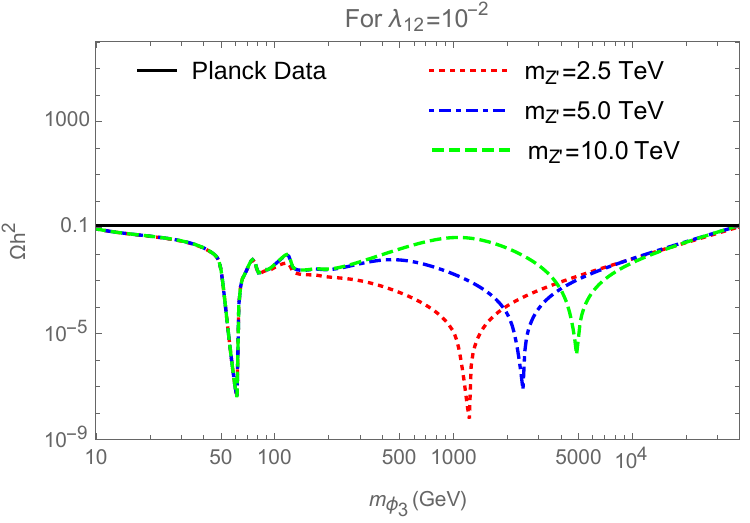}\includegraphics[scale=0.6]{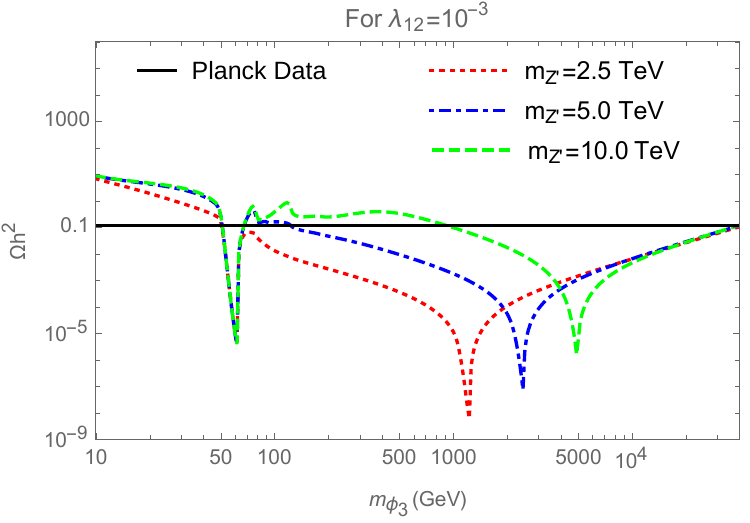}
\caption{Relic abundance as a function of dark matter mass for various $Z^\prime$ masses in the multi-component dark matter scenario.}
\label{fig:relic_vs_mdm_var_mZp}
\end{figure}

Next, we turn to the total relic abundance as a function of DM mass in Fig.~\ref{fig:relic_vs_mdm_var_mZp} but this time for various values of $Z^\prime$ mass and two benchmark values of $\lambda_{12}$ to complement our results in Fig.~\ref{fig:relic_final}. For the higher DM mass range not much difference can be noticed in this logarithmic scale plot for various $m_{Z^\prime}$ values apart from the shift in the Breit-Wigner resonance. However, in the lower DM mass range there is strong competition from the Higgs mediated channels. A heavier $Z^\prime$ mediated process generates less cross section compared to the Higgs mediated processes in low DM mass region. However, a lighter $Z^\prime$ mediated process can produce comparable cross section to the higgs mediated ones and thus contribute to the relic significantly even in the low DM mass range as evident in Fig.~\ref{fig:relic_vs_mdm_var_mZp} (right).

\subsection{Direct Detection Constraints}
\label{subsec:dirdet}
Direct detection experiments like Xenon1T \cite{Aprile:2018dbl,XENON:2022avm}, PandaX-4T \cite{PandaX-4T:2021bab} and LUX-ZEPLIN (LZ) \cite{LZ:2022ufs} constrain the parameter space by non-observation as they provide upper limits on the scattering cross-section of the DM candidates with the detector nuclei in the non-relativistic limit. The expression\footnote{We have assumed that the quark contribution to each of the nucleons are approximately equal.} for spin independent scattering cross section of $\phi_0$ with the nucleons of the detector material is given by \cite{Basak:2021tnj}
\begin{equation}
\sigma_{N,0}^{SI}=\frac{\mu_0^2}{4\pi m_{\phi_0}^2}\bigg[\frac{\lambda_{h\phi_0\phi_0}}{m_h^2}\bigg(\sum_{q=u,d,s}\lambda_{hqq}\frac{m_N}{m_q}f_{Tq}^N+\frac{2}{27}f_{TG}^N\sum_{q=c,b,t}\lambda_{hqq}\frac{m_N}{m_q}\bigg)\bigg]^2,
\label{eq:dir_detect_0h}
\end{equation}
where, $\mu_0=\frac{m_{\phi_0}m_N}{m_{\phi_0}+m_N}$. The matrix elements ($f^N_{Ti}$) are given in reference \cite{Ellis:2000ds}.

\begin{figure}[h!]
\centering
\includegraphics[scale=0.65]{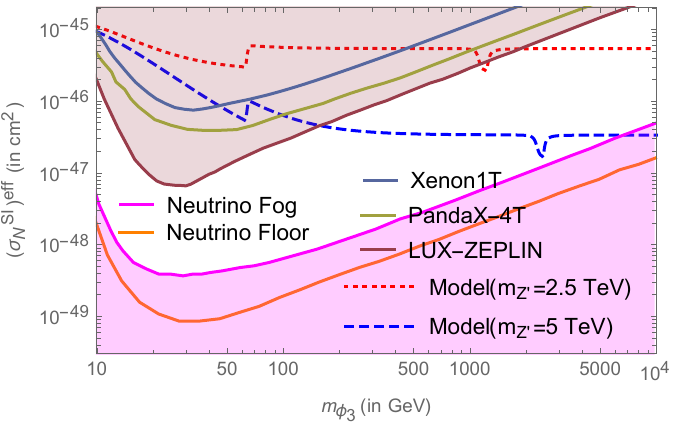}\includegraphics[scale=0.65]{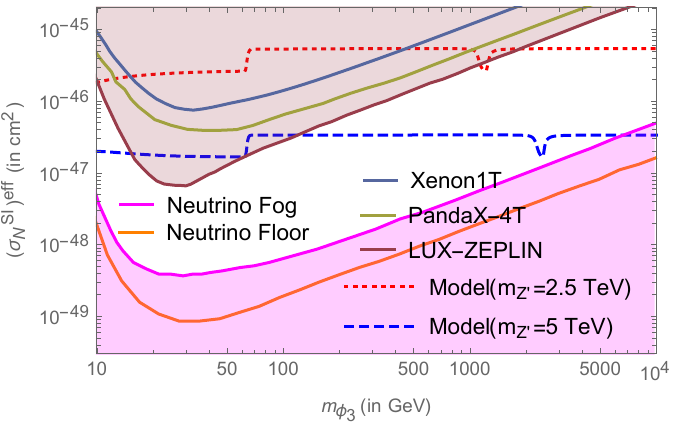}
\caption{Constraints from direct detection experiments on the effective cross-section of dark matter candidate $\phi_0$ for benchmark values $\lambda_{12}=10^{-2}$(left) and $\lambda_{12}=10^{-3}$(right). The constraints from Xenon1T(2022), PandaX-4T (2021), LUX-ZEPLIN (2022) and also the Neutrino Floor have been shown in the figures. Also, note that the mass degeneracy ($m_{\phi_0}=m_{\phi_3}$ at the limit ($\lambda_{23}+\tilde\lambda_{23})\rightarrow0$ following Eqn.~\ref{equ:sca_mass}) assumption makes the two masses interchangeable.}
\label{fig:dirdetect}
\end{figure}

In addition to the Higgs mediated $\phi_0N\rightarrow \phi_0N$ scattering process mentioned above, this model also exhibits a $Z^\prime$ mediated $\phi_0N\rightarrow \phi_3N$ process. In the mass degeneracy limit of the DM candidates, both the contributions are to be added together to arrive at the combined cross-section. The relevant couplings $\lambda_{\phi_0\phi_3Z^\prime}=-\lambda_{\phi_3\phi_0Z^\prime}$ and $\lambda_{Z^\prime\bar u_Lu_L}=-\lambda_{Z^\prime\bar d_Ld_L}$ are given in Tables \ref{tab:BSM_scalar_derivative_gauge_cubic} and \ref{tab:BSM_gauge_quark_cubic}. 
In the multicomponent dark matter scenario, the quantity of relevance to realise the constraints from direct detection experiments by non-observation is given by the following effective cross-section\footnote{for a detailed study with non-degenerate masses of the DM candidates and the computation of total DM scattering rate see Ref.~\cite{Profumo:2009tb}.} \cite{Bhattacharya:2016ysw,Herrero-Garcia:2017vrl,Bhattacharya:2019tqq},
\begin{equation}
(\sigma_{N,i}^{SI})^{eff}=\frac{\Omega_i}{\Omega_{tot}}\sigma_{N,i}^{SI},
\label{eq:eff_sigma}
\end{equation}
where the DM-nucleon scattering cross-section is modified with each DM candidate's respective contribution to the relic abundance and $\sigma_{N,i}^{SI}$ is the summation of the contributions from both diagrams $\phi_iN\rightarrow \phi_iN$ and $\phi_iN\rightarrow \phi_jN$. Figure \ref{fig:dirdetect} describes the experimental constraints coming from Xenon1T, PandaX-4T and LZ experiments on the effective cross-section of the DM candidate $\phi_0$ as a function of DM mass where LZ is clearly the most restricting of them all. A similar set of constraints can be obtained for the $\phi_3$ as well following the same procedure.

\begin{figure}[h!]
\centering
\includegraphics[scale=0.65]{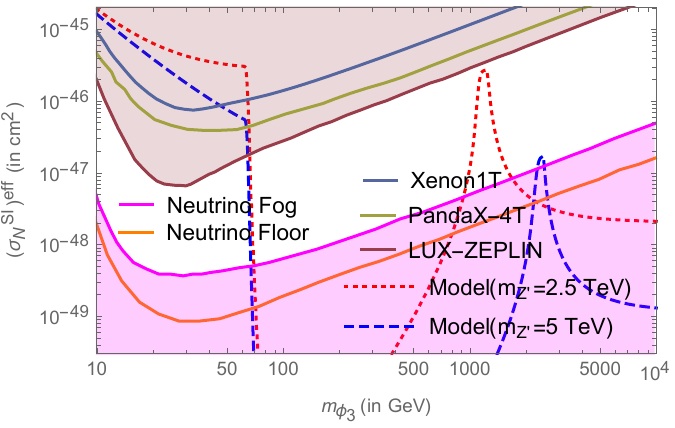}\includegraphics[scale=0.65]{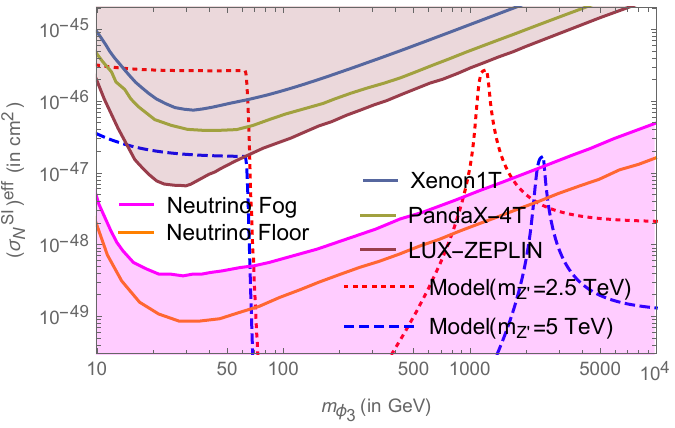}
\caption{Same as figure \ref{fig:dirdetect} but for $\phi_3$.}
\label{fig:dirdetect_phi_3}
\end{figure}

It is evident from Fig.~\ref{fig:dirdetect} that for a $Z^\prime$ mass of 2.5 TeV (motivated by the lower bound from Fig.~\ref{fig:Zpbound}), much of the DM mass parameter space is already excluded upto around 2 TeV for both the benchmark values of coupling parameter $\lambda_{12}=10^{-2}$ and $10^{-3}$. Keeping every other benchmark value same and modifying only the $Z^\prime$ mass to 5 TeV brings the DM mass exclusion region down to $\sim100-200$ GeV range and very interestingly opens up some parameter space in the lower DM mass region of around $10-15$ GeV for $\lambda_{12}=10^{-3}$. A more careful look at the figures makes it apparent that for $\lambda_{12}=10^{-2}$, the Higgs mediated diagram plays an important role upto the DM masses around the weak scale and beyond that the heavy gauge boson mediated diagram takes over. However, the contribution of the Higgs mediated diagram remains subdominant to the one mediated by the heavy gauge boson throughout the DM mass range for $\lambda_{12}=10^{-3}$. As expected, the cross section is more suppressed by a higher value of the $Z^\prime$ mass compared to a lower value. The slightly raised plateau like shape of the curves in the mid-range of DM masses depicted here are the result of the fractional contributions of the individual DM components to the relic. One also has to consider the neutrino floor \cite{Billard:2021uyg} and the neutrino fog \cite{OHare:2021utq} in this regard as for cross-sections too small the DM scattering will be difficult to tell apart due to the presence of neutrino scattering. Similar results are expected for $\phi_3$ upto the relic fraction as is realized in Fig.~\ref{fig:dirdetect_phi_3}. The rise (dip) in the curves are indicative of the substantial (negligible) contribution of $\phi_3$ to the relic in that DM mass region.
Even though in this work we have worked under the mass degeneracy limit of the DM candidates (i.e. $m_{\phi_0}=m_{\phi_3}$), there is also in principle a scope for inelastic scattering between the two DM components by softly breaking their mass degeneracy. However, that would lead to some problems with the stability of the DM components and this multicomponent DM scenario will effectively reduce to a single component scenario as discussed in Sec.~\ref{subsec:dm_can}.


\section{Conclusion}
\label{sec:conclusion}
In this work we have extended the SM with an SU(2) gauge sector and the scalar sector with a bidoublet and a non-linear sigma field. The neutral components of the $Z_2$ protected bidoublet are suitable DM candidates. In the DM mass range starting from a few GeV upto around half of the Higgs mass, the constraint from invisible decay of higgs becomes important and restricts the most relevant scalar sector coupling parameter $\lambda_{12}$ with an upper bound of $\mathcal{O}(10^{-3}-10^{-2})$. However, the sensitivity of direct detection experiments provide a much more stringent bound for a typical scalar extension under the WIMP scenario. In many scalar extensions, one has to resort to the Breit-Wigner resonance to account for the DM relic abundance while simultaneously satisfying the direct detection constraints. In our multicomponent DM model the extended gauge sector mitigates those strict constraints to a great extent. The gauge sector couplings provide additional channels for the DM annihilation and co-annihilation thus increasing the velocity averaged cross-section and reducing the relic to avoid overabundance. In the similar coupling range this model can also be constrained with the direct detection effective cross-section values from the experimental observations. Furthermore, we found that a $Z^\prime$ mediated scattering diagram can dominate a scalar mediated diagram in a sizeable

\acknowledgments
 BC acknowledges support from the Department of Science and Technology, India, under Grant CRG/2020/004171. KL thanks \'Oscar Zapata for helpful comments. AS thanks Tarak Nath Maity for helpful discussions.
\appendix

\section{Perturbativity}
\label{ap:per}

\begin{figure}[h!]
\centering
\includegraphics[scale=0.4]{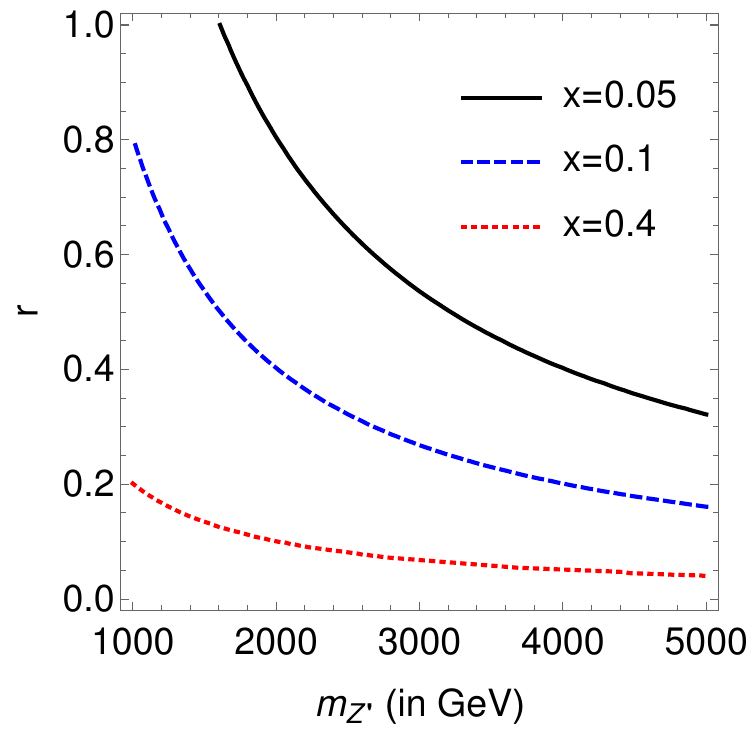}
\caption{The variation of $x$ in $M_{Z^{'}}$ versus $r$ plane. For details please see the text.}
\label{fig:mZpvsx}
\end{figure}

The perturbativity of a model typically means that the loop level contributions towards the scattering amplitude should be lower than the tree level contribution. It must be ensured that the available couplings in the model are sufficiently small such that one can employ perturbation theory reliably to calculate the $s$-matrix.
In this model the cubic vertices between the $Z^\prime$ boson and $\phi_{0}\phi_{3}$ or $\phi^{+}\phi^{-}$ scalars are proportional to $\frac{1}{x}$. As a result, for a very small values of $x$ the decay width $\Gamma(Z^{'} \to \phi_{0}\phi_{3}/\phi^{+}\phi^{-})$ becomes significantly large. For those values of $x$, the perturbative treatment is not applicable. From Eq.~\ref{eq:xdef} one can see $x = \sqrt{\frac{M^{2}_{W}}{r^{2}M^{2}_{Z^{'}}}}$. This relation suggest that the parameter $x$ is dependent on two variables $r$ and $M_{Z^{'}}$. In fig.~\ref{fig:mZpvsx}, we show the variation of the parameter $x$ in $M_{Z^{'}}$ vs $r$ plane. 

\begin{figure}[h!]
\centering
\includegraphics[scale=0.4]{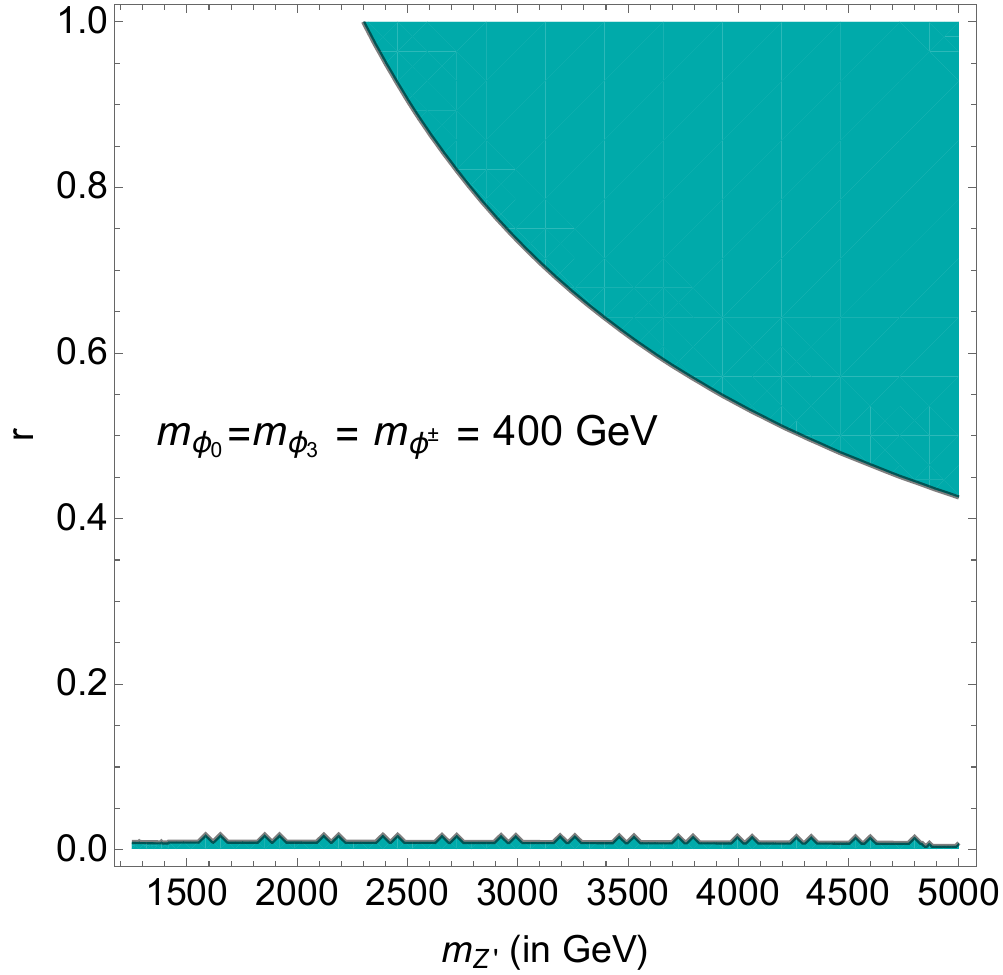}
\caption{The perturbativity limit is well respected in all regions of interest. The dark cyan shaded portion represents the disallowed region of parameter space.}
\label{fig:perturb}
\end{figure}

The black line, blue dashed and red dotted lines signify the contours of $x$ equals to 0.05, 0.1 and 0.5 respectively. From the plot one can notice for the $r \to 0$ the value of $x$ enhances for the entire range of $M_{Z^{'}}$. With these numbers in hand one can evaluate $\frac{\Gamma_{Z^{'}}}{M_{Z^{'}}}$ for the $r$ ranging from $10^{-1}$ to 1 and $M_{Z^{'}}$ ranging from 1.25 TeV to 5 TeV. In fig.~\ref{fig:perturb}, the allowed region is $\frac{\Gamma_{Z^{'}}}{M_{Z^{'}}}$ < 1, where one can safely consider perturbativity.  
\FloatBarrier

\section{Couplings}
\label{ap:coup}

The couplings are arranged in a tabular form as the following (where $\theta$ is the Weinberg angle):

\begin{table}[h!]
\begin{center}
\begin{tabular}{|c|c|}
\hline
$\lambda_{abc}$& Couplings \\
\hline
$hhh$&$\lambda_1v$\\
\hline
$h\phi_0\phi_0$&$\frac{\lambda_{12}v}{2}$\\
\hline
$h\phi_3\phi_3$&$\frac{\lambda_{12}v}{2}$\\
\hline
$h\phi^+\phi^-$&$\lambda_{12}v$\\
\hline
\end{tabular}
\caption{Cubic scalar couplings.}
\label{tab:scalar_cubic}
\end{center}
\end{table}

\begin{table}[h!]
\begin{center}
\begin{tabular}{|c|c|}
\hline
$\lambda_{abcd}$& Couplings \\
\hline
$hhhh$&$\frac{\lambda_1}{4}$\\
\hline
$hh\phi_0\phi_0$&$\frac{\lambda_{12}}{4}$\\
\hline
$hh\phi_3\phi_3$&$\frac{\lambda_{12}}{4}$\\
\hline
$hh\phi^+\phi^-$&$\frac{\lambda_{12}}{2}$\\
\hline
$\phi_0\phi_0\phi_0\phi_0$&$\frac{\lambda_2}{4}$\\
\hline
$\phi_3\phi_3\phi_3\phi_3$&$\frac{\lambda_2}{4}$\\
\hline
$\phi^+\phi^+\phi^-\phi^-$&$\lambda_2$\\
\hline
$\phi_0\phi_0\phi_3\phi_3$&$\frac{\lambda_2}{2}$\\
\hline
$\phi^0\phi^0\phi^+\phi^-$&$\lambda_2$\\
\hline
$\phi^3\phi^3\phi^+\phi^-$&$\lambda_2$\\
\hline
\end{tabular}
\caption{Quartic scalar couplings.}
\label{tab:scalar_quartic}
\end{center}
\end{table}

\begin{table}[h!]
\begin{center}
\begin{tabular}{|c|c|}
\hline
$\lambda_{abc}$& Couplings \\
\hline
$hZZ$&$\frac{e^2v}{4}\sec^2\theta(\cosec^2\theta+x^2)$\\
\hline
$hZ^\prime Z^\prime$&$\frac{e^2v}{4}x^2\cosec^2\theta$\\
\hline
$hZZ^\prime$&$-\frac{e^2v}{2}x\cosec^2\theta\sec\theta$\\
\hline
$hW^+W^-$&$\frac{e^2v}{2}(\cosec^2\theta+x^2)$\\
\hline
$hW^{\prime+}W^{\prime-}$&$\frac{e^2v}{2}x^2\cosec^2\theta$\\
\hline
$hW^{\prime+}W^-$&$-\frac{e^2v}{2}x\cosec^2\theta$\\
\hline
\end{tabular}
\caption{Cubic higgs-gauge couplings.}
\label{tab:higgs_gauge_cubic}
\end{center}
\end{table}

\begin{table}[h!]
\begin{center}
\begin{tabular}{|c|c|c|}
\hline
$\lambda_{abcd}$& Couplings \\
\hline
$hhZZ$&$\frac{e^2}{8}\sec^2\theta(\cosec^2\theta+x^2)$\\
\hline
$hhZ^\prime Z^\prime$&$\frac{e^2}{8}x^2\cosec^2\theta$\\
\hline
$hhZZ^\prime$&$-\frac{e^2}{4}x\cosec^2\theta\sec\theta$\\
\hline
$hhW^+W^-$&$\frac{e^2}{4}(\cosec^2\theta+x^2)$\\
\hline
$hhW^{\prime+}W^{\prime-}$&$\frac{e^2}{4}x^2\cosec^2\theta$\\
\hline
$hhW^{\prime+}W^-$&$-\frac{e^2}{4}x\cosec^2\theta$\\
\hline
\end{tabular}
\caption{Quartic higgs-gauge couplings.}
\label{tab:higgs_gauge_quartic}
\end{center}
\end{table}

\begin{table}[h!]
\begin{center}
\begin{tabular}{|c|c|}
\hline
$\lambda_{abcd}$& Couplings\\
\hline
$\phi_0\phi_0Z^\prime Z^\prime$&$\frac{e^2\cosec^2\theta}{8x^2}+\frac{e^2}{8}(3+2\cot^2\theta)+\frac{e^2}{8}(3+\cot^2\theta)x^2$\\
\hline
$\phi_0\phi_0W^{\prime+}W^{\prime-}$&$\frac{e^2\cosec^2\theta}{4x^2}+\frac{e^2}{4}(3+2\cot^2\theta)+\frac{e^2}{4}(3+\cot^2\theta)x^2$\\
\hline
$\phi_3\phi_3W^+W^-$&$e^2(\cosec^2\theta+x^2)$\\
\hline
$\phi_3\phi_3W^{\prime+}W^{\prime-}$&$\frac{e^2}{4x^2}\cosec^2\theta-\frac{e^2}{4}(\cosec^2\theta+\cot^2\theta)+\frac{e^2}{4}(\cot^2\theta-1)x^2$\\
\hline
$\phi_3\phi_3W^{\prime+}W^-$&$\frac{e^2}{2x}\cosec^2\theta-\frac{e^2}{2}(\cot^2\theta)x$\\
\hline
$\phi_0\phi_3W^{\prime+}W^-$&$\frac{ie^2}{2x}\cosec^2\theta+\frac{ie^2}{2}(\cosec^2\theta+1)x$\\
\hline
\end{tabular}
\caption{Quartic gauge couplings for DM annihilation.}
\label{tab:DM_gauge_quartic}
\end{center}
\end{table}

\begin{table}[h!]
\begin{center}
\begin{tabular}{|c|c|}
\hline
$\lambda_{abcd}$& Couplings \\
\hline
$\phi^+\phi_0W^-Z^\prime$&$-\frac{ie^2}{2x}\cosec^2\theta-\frac{ie^2}{2}(\cosec^2\theta+1)x$\\
\hline
$\phi^+\phi_3W^-A$&$-e^2\cosec\theta-e^2(\sin\theta)x^2$\\
\hline
$\phi^+\phi_3W^-Z$&$-e^2\cos\theta(\cosec^2\theta+x^2)$\\
\hline
$\phi^+\phi_3W^-Z^\prime$&$-\frac{e^2}{2x}\cosec^2\theta+\frac{e^2}{2}(\cot^2\theta)x$\\
\hline
$\phi^+\phi_0W^{\prime-}A$&$\frac{ie^2}{2x}\cosec\theta+\frac{ie^2}{2}(\cosec\theta+\sin\theta)$\\
\hline
$\phi^+\phi_0W^{\prime-}Z$&$\frac{ie^2}{2x}\cot\theta\cosec\theta+\frac{ie^2}{2}\cos\theta(2+\cot^2\theta)x$\\
\hline
$\phi^+\phi_3W^{\prime-}A$&$-\frac{e^2}{2x}\cosec\theta+\frac{e^2}{2}(\cos\theta\cot\theta)x$\\
\hline
$\phi^+\phi_3W^{\prime-}Z$&$-\frac{e^2}{2x}\cot\theta\cosec\theta+\frac{e^2}{2}(\cos\theta\cot^2\theta)x$\\
\hline
$\phi^+\phi_3W^{\prime-}Z^\prime$&$e^2(\cosec^2\theta+x^2)$\\
\hline
\end{tabular}
\caption{Quartic couplings for charged scalar decay.}
\label{tab:charged_scalar_DM_gauge_quartic}
\end{center}
\end{table}

\begin{table}[h!]
\begin{center}
\begin{tabular}{|c|c|}
\hline
$\lambda_{abcd}$& Couplings \\
\hline
$\phi^+\phi^-AA$&$e^2+e^2\sin^2\theta x^2$\\
\hline
$\phi^+\phi^-ZZ$&$e^2\cot^2\theta+e^2\cos^2\theta x^2$\\
\hline
$\phi^+\phi^-Z^\prime Z^\prime$&$\frac{e^2}{4x^2}\cosec^2\theta-\frac{e^2}{4}(1+2\cot^2\theta)-\frac{e^2}{4}(1-\cot^2\theta)x^2$\\
\hline
$\phi^+\phi^-AZ$&$2e^2\cot\theta+2e^2(\sin\theta\cos\theta)x^2$\\
\hline
$\phi^+\phi^-AZ^\prime$&$\frac{e^2}{x}\cosec\theta-e^2(\cos\theta\cot\theta)x$\\
\hline
$\phi^+\phi^-ZZ^\prime$&$\frac{e^2}{x}\cot\theta\cosec\theta-e^2(\cos\theta\cot^2\theta)x$\\
\hline
$\phi^+\phi^-W^+W^-$&$e^2(\cosec^2\theta+x^2)$\\
\hline
$\phi^+\phi^-W^{\prime+}W^{\prime-}$&$\frac{e^2}{2x^2}\cosec^2\theta+\frac{e^2}{2}+\frac{e^2}{2}(\cosec^2\theta)x^2$\\
\hline
$\phi^+\phi^-W^{\prime+}W^-$&$\frac{e^2}{2x}\cosec^2\theta-\frac{e^2}{2}(\cot^2\theta)x$\\
\hline
\end{tabular}
\caption{Quartic couplings for charged scalar annihilation.}
\label{tab:charged_scalar_gauge_quartic}
\end{center}
\end{table}

\begin{table}[h!]
\begin{center}
\begin{tabular}{|c|c|}
\hline
$\lambda_{abcd}$& Couplings\\
\hline
$\phi^+\phi^+W^-W^-$&$-\frac{e^2}{2}(\cosec^2\theta+x^2)$\\
\hline
$\phi^+\phi^+W^{\prime-}W^{\prime-}$&$\frac{e^2}{2}(\cosec^2\theta+x^2)$\\
\hline
$\phi^+\phi^+W^{\prime-}W^-$&$-\frac{e^2}{2x}\cosec^2\theta+\frac{e^2}{2}(\cot^2\theta)x$\\
\hline
\end{tabular}
\caption{A few more couplings between the scalar and gauge sectors.}
\label{tab:charged_gauge_quartic}
\end{center}
\end{table}

\begin{table}[h!]
\begin{center}
\begin{tabular}{|c|c|c|}
\hline
$\lambda_{abcd}$& Couplings \\
\hline
$(\partial\phi_0)\phi_3Z^\prime$&$-\frac{e}{2x}\cosec\theta-\frac{e}{4\sec\theta}(2\cot\theta+3\tan\theta)x$\\
\hline
$(\partial\phi_0)\phi^+W^{\prime-}$&$-\frac{e}{2x}\cosec\theta-\frac{e}{4\sec\theta}(2\cot\theta+3\tan\theta)x$\\
\hline
$(\partial\phi_3)\phi_0Z^\prime$&$\frac{e}{2x}\cosec\theta+\frac{e}{4\sec\theta}(2\cot\theta+3\tan\theta)x$\\
\hline
$(\partial\phi_3)\phi^+W^-$&$-ie\cosec\theta-\frac{ie}{2}(\sin\theta)x^2$\\
\hline
$(\partial\phi_3)\phi^+W^{\prime-}$&$-\frac{ie}{2x}\cosec\theta+\frac{ie}{4\sec\theta}(2\cot\theta+\tan\theta)x$\\
\hline
$(\partial\phi^+)\phi_0W^{\prime-}$&$\frac{e}{2x}\cosec\theta+\frac{e}{4\sec\theta}(2\cot\theta+3\tan\theta)x$\\
\hline
$(\partial\phi^+)\phi_3W^-$&$ie\cosec\theta+\frac{ie}{2}(\sin\theta)x^2$\\
\hline
$(\partial\phi^+)\phi_3W^{\prime-}$&$\frac{ie}{2x}\cosec\theta-\frac{ie}{4\sec\theta}(2\cot\theta+\tan\theta)x$\\
\hline
$(\partial\phi^+)\phi^-A$&$-ie-\frac{ie}{2}(\sin^2\theta)x^2$\\
\hline
$(\partial\phi^+)\phi^-Z$&$-ie\cot\theta-\frac{ie}{2}(\sin\theta\cos\theta)x^2$\\
\hline
$(\partial\phi^+)\phi^-Z^\prime$&$-\frac{ie}{2x}\cosec\theta+\frac{ie}{4\sec\theta}(2\cot\theta+\tan\theta)x$\\

\hline
\end{tabular}
\caption{Cubic Couplings of gauge fields and BSM scalars involving derivatives.}
\label{tab:BSM_scalar_derivative_gauge_cubic}
\end{center}
\end{table}

\FloatBarrier

\begin{table}[h!]
\begin{center}
\begin{tabular}{|c|c|c|}
\hline
$\lambda_{abc}$& Couplings\\
\hline
$Z\bar{u}_Lu_L$&$\frac{g_1\cos\theta}{2}-g_2\tilde Y_u\sin\theta$\\
\hline
$Z^\prime\bar{u}_Lu_L$&$-\frac{g_1x}{2}$\\
\hline
$A\bar{u}_Lu_L$&$\frac{g_1\sin\theta}{2}+g_2\cos\theta\tilde Y_u$\\
\hline
$Z\bar{d}_Ld_L$&$-\frac{g_1\cos\theta}{2}-g_2\tilde Y_d\sin\theta$\\
\hline
$Z^\prime\bar{d}_Ld_L$&$\frac{g_1x}{2}$\\
\hline
$A\bar{d}_Ld_L$&$-\frac{g_1\sin\theta}{2}+g_2\cos\theta\tilde Y_d$\\
\hline
$Z\bar{u}_Ru_R$&$-g_2\sin\theta\tilde Y_u$\\
\hline
$Z^\prime\bar{u}_Ru_R$&$0$\\
\hline
$A\bar{u}_Ru_R$&$g_2\cos\theta\tilde Y_u$\\
\hline
$Z\bar{d}_Rd_R$&$-g_2\sin\theta\tilde Y_d$\\
\hline
$Z^\prime\bar{d}_Rd_R$&$0$\\
\hline
$A\bar{d}_Rd_R$&$g_2\cos\theta\tilde Y_d$\\
\hline
$Wud$&$\frac{g_1}{\sqrt{2}}V_{CKM}$\\
\hline
$W^\prime ud$&$-\frac{g_1x}{\sqrt{2}}V_{CKM}$\\

\hline
\end{tabular}
\caption{Cubic Couplings of gauge fields with quarks.}
\label{tab:BSM_gauge_quark_cubic}
\end{center}
\end{table}
\FloatBarrier

\begin{table}[h!]
\begin{center}
\begin{tabular}{|c|c|c|}
\hline
$\lambda_{abc}$& Couplings\\
\hline
$Z\bar{e}_Re_R$&$-g_2\sin\theta\tilde Y_e$\\
\hline
$Z^\prime\bar{e}_Re_R$&$0$\\
\hline
$A\bar{e}_Re_R$&$g_2\cos\theta\tilde Y_e$\\
\hline
$Z\bar{\nu}_L\nu_L$&$\frac{g_1\cos\theta}{2}-g_2\sin\theta\tilde Y_l$\\
\hline
$Z^\prime\bar{\nu}_L\nu_L$&$-\frac{g_1x}{2}$\\
\hline
$A\bar{\nu}_L\nu_L$&$\frac{g_1\sin\theta}{2}+g_2\cos\theta\tilde Y_l$\\
\hline
$W\bar{e}_L\nu_L$&$\frac{g_1}{\sqrt{2}}$\\
\hline
$W^\prime\bar{e}_L\nu_L$&$-\frac{g_1x}{\sqrt{2}}$\\

\hline
\end{tabular}
\caption{Cubic Couplings of gauge fields with leptons.}
\label{tab:BSM_gauge_lepton_cubic}
\end{center}
\end{table}

\begin{table}[h!]
\begin{center}
\begin{tabular}{|c|c|c|}
\hline
$\lambda_{abc}$ & Couplings \\
\hline
$WW\gamma$ & $e\left(1 + x^{2}\right)$ \\
\hline
$W^{'}W^{'}\gamma$ & $e\left(1 + x^{2}\right)$ \\
\hline
$WW^{'}\gamma$ & 0 \\
\hline
$WWZ$ & $e\cot\theta_{w}\left(1 + x^{2}\right)$ \\
\hline
$W^{'}W^{'}Z^{'}$ & $\frac{e}{x\sin\theta_{w}}\left(1 - x^{4}\right)$ \\
\hline
$WWZ^{'}$ & $-\frac{ex}{\sin\theta_{w}}\left(1 + x\right)$ \\
\hline
\end{tabular}
\caption{Relevant couplings of gauge bosons among themselves.}
\label{tab:gauge}
\end{center}
\end{table}
\FloatBarrier

\section{$h \to \gamma \gamma$ Decay Width}
\label{ap:h_gamma_gamma}
\begin{figure}
\centering
\includegraphics[scale=0.6]{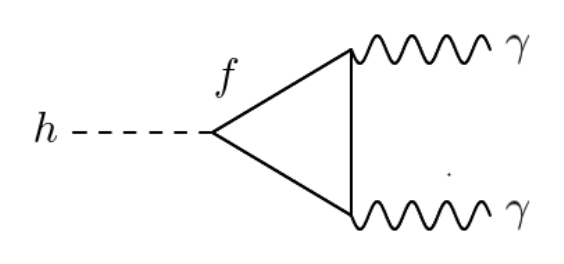}
\includegraphics[scale=0.6]{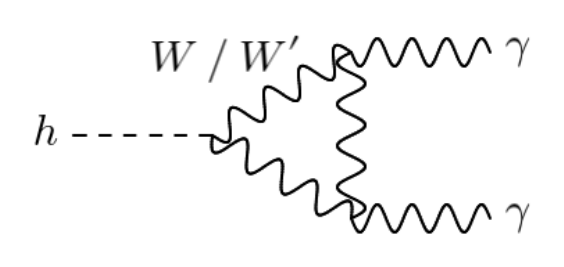}
\includegraphics[scale=0.6]{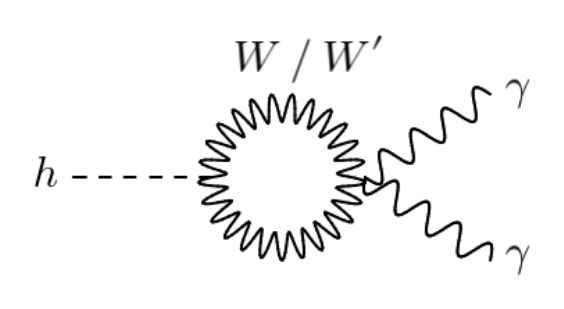}
\includegraphics[scale=0.6]{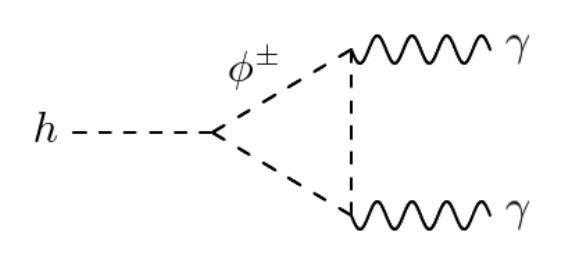}
\includegraphics[scale=0.6]{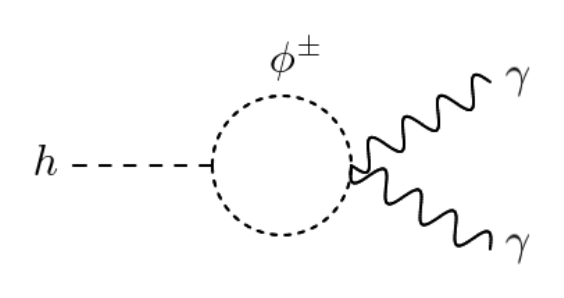}
\caption{All possible loop mediated processes that can contribute to $h \to \gamma \gamma$ decay mode.}
\label{fig:hLoops}
\end{figure}

The CP-even Higgs boson can decay into di-photon channel through loop mediated processes. In case of SM, the $W$ boson loop and different fermion loops participate in this process. For the present model, in addition to these loop diagrams the heavy charged gauge boson $W^{'}$ and the charged scalar $\phi^{\pm}$ can also engage in the loop induced processes. In Fig.~\ref{fig:hLoops} we show the all the Feynman diagrams that can contribute to these process. Ref.~\cite{Gunion:1989we} have calculated this decay width for MSSM scenario and adopting their formalism one can reproduce the width for the current model.  The analytic expression for this decay width is given in Eq.~\ref{eq:h2AA} 
\begin{equation}
\Gamma\left(h \to \gamma \gamma\right) = \frac{\alpha^{2}g^{2}}{1024\pi^{3}}\frac{m^{3}_{h}}{m^{2}_{W}}\left|\sum_{i}\mathcal{I}^{i}_{h}\right|^{2}
\label{eq:h2AA}
\end{equation}

\noindent
where $\mathcal{I}^{i}_{h}$ represent the individual loop contribution from different particles that flowing in the loop. In Eq.~\ref{eq:Ihi}, we present the explicit expression for $\mathcal{I}^{i}_{h}$. 
\begin{equation}
\begin{split}
\mathcal{I}^{f}_{h} &= N_{c}Q^{2}_{f}R^{h}_{f}F^{h}_{\frac{1}{2}}\left(\tau_{f}\right) \\
\mathcal{I}^{W/W^{'}}_{h} &= R^{h}_{W/W^{'}}F_{1}\left(\tau_{W/W^{'}}\right) \\
\mathcal{I}^{\phi^{\pm}}_{h} &= R^{h}_{\phi^{\pm}}\frac{M^{2}_{W}}{M^{2}_{\phi^{\pm}}}F_{0}\left(\tau_{\phi^{\pm}}\right)
\end{split}
\label{eq:Ihi}
\end{equation}
Here $N_{c}$ and $Q_{f}$ represents the color factor and electromagnetic charges for each SM fermions. The $R^{h}_{i}$s are the relative coupling strength of the three point vertices $\lambda_{hii}$ (where $i$ can vary based on the particle flowing in the loop) \emph{w.r.t} corresponding SM coupling. For the present model $R^{h}_{f} = 1$ as the fermion sector is identical to the SM. In case of gauge bosons and the charged scalar the relative coupling strength would be $R^{h}_{W} = \left(1 + x^{2}\sin^{2}\theta_{W}\right)$, $R^{h}_{W^{'}} = x^{2}$ and $R^{h}_{\phi^{\pm}} = \frac{2\lambda_{ab}}{g^{2}}$ respectively. To get the sense of these values one should the check the corresponding vertices that are present in appendix \ref{ap:coup}. The $F_{j}\left(\tau_{i}\right)$ are the form factors corresponding to the gauge, fermion and scalar loops which can take following form. 
\begin{equation}
\begin{split}
F_{0} & = \tau\left[1 - \tau f\left(\tau\right)\right] \\
F_{\frac{1}{2}} & = -2\tau\left[1 + \left(1 - \tau\right)f\left(\tau\right)\right] \\
F_{1} & = 2 + 3\tau + 3\tau\left(2 - \tau\right)f\left(\tau\right)
\end{split}
\end{equation}
where $\tau_{i} = \frac{4m^{2}_{i}}{m^{2}_{h}}$ and depending on the value of $\tau$ the $f\left(\tau\right)$ can take following form 
\begin{equation}
\begin{split}
f\left(\tau\right) & = \left[\sin^{-1}\left(\sqrt{\frac{1}{\tau}}\right) \right]^{2} ~~~ \text{if}~~~ \tau \geq 1 \\
\text{or} \\
f\left(\tau\right) & = -\frac{1}{4}\left[\text{ln}\left(\frac{1 + \sqrt{1 - \tau}}{1 + \sqrt{1 - \tau}}\right) - i\pi\right]^{2} ~~~ \text{if} ~~\tau \leq 1
\end{split}
\end{equation}

\section{Boltzmann Equations}
\label{ap:beq}

\begin{figure}
\centering
\includegraphics[scale=0.5]{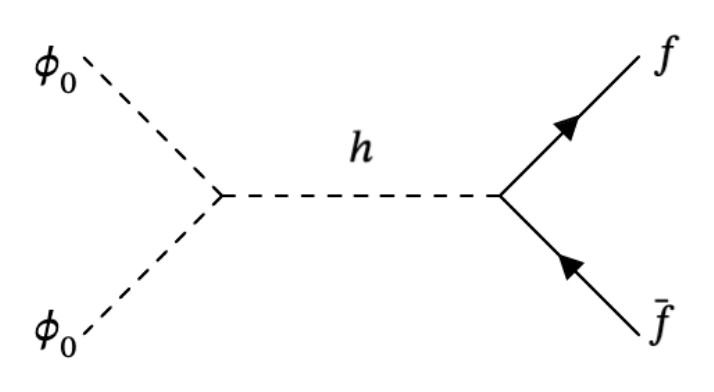}
\includegraphics[scale=0.5]{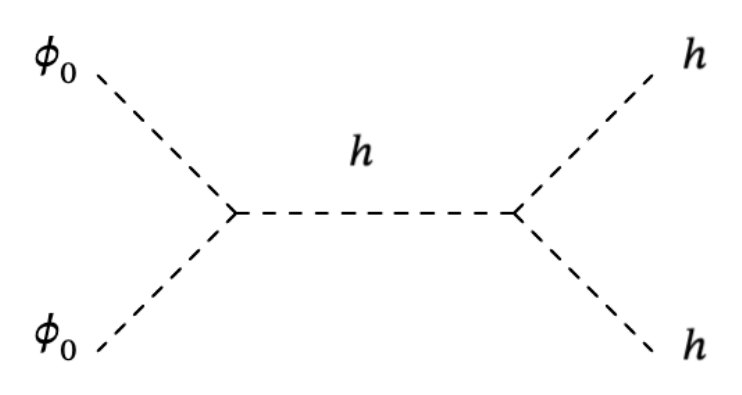}
\includegraphics[scale=0.5]{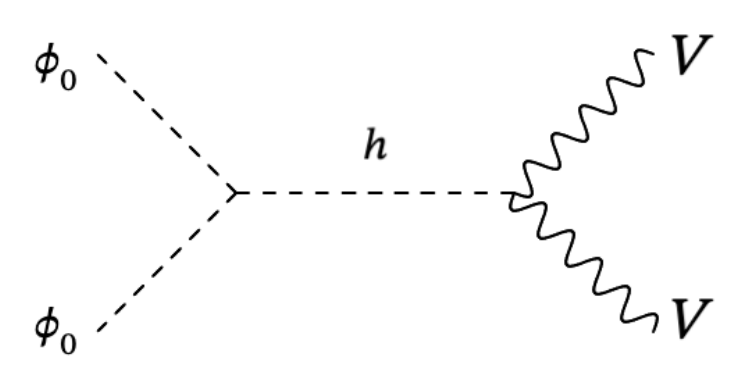}
\includegraphics[scale=0.5]{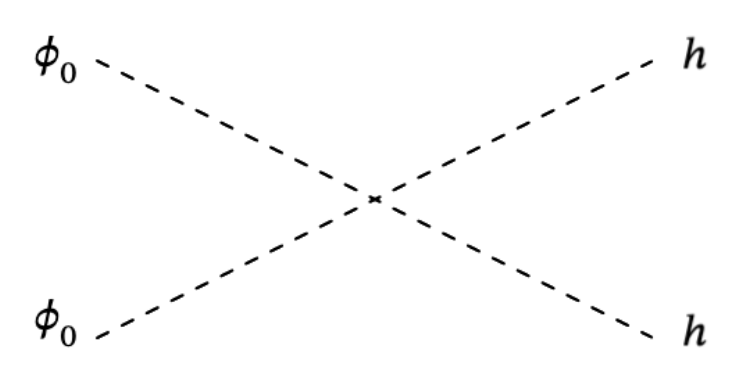}
\includegraphics[scale=0.6]{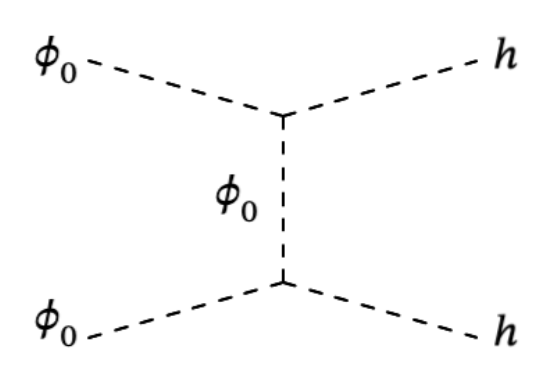}
\caption{Relevant $\phi_0$ annihilation channels for relic abundance calculation.}
\label{fig:phi_0_anni}
\end{figure}

\begin{figure}
\centering
\includegraphics[scale=0.5]{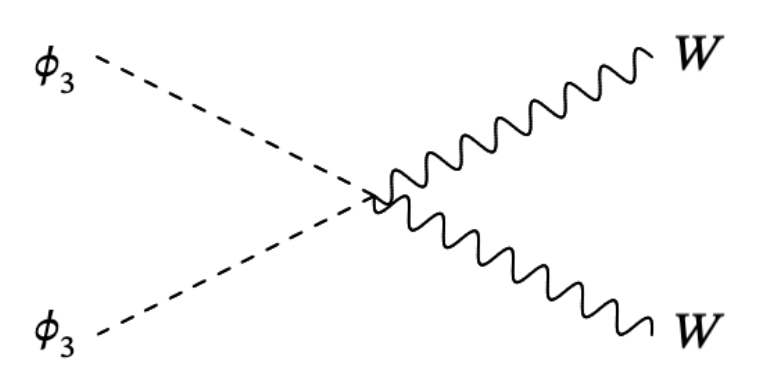}
\includegraphics[scale=0.55]{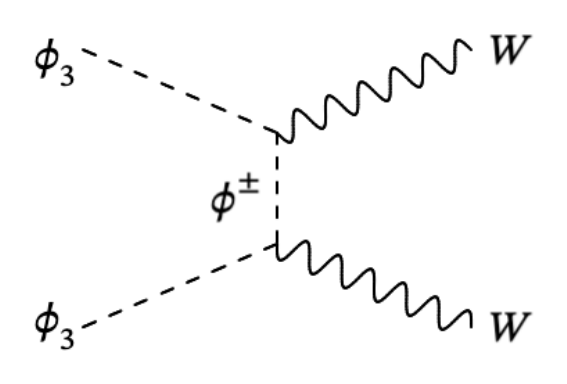}
\caption{The extra $\phi_3$ annihilation channels for relic abundance calculation compared to $\phi_0$ annihilation channels.}
\label{fig:phi_3_anni}
\end{figure}

\begin{figure}
\centering
\includegraphics[scale=0.5]{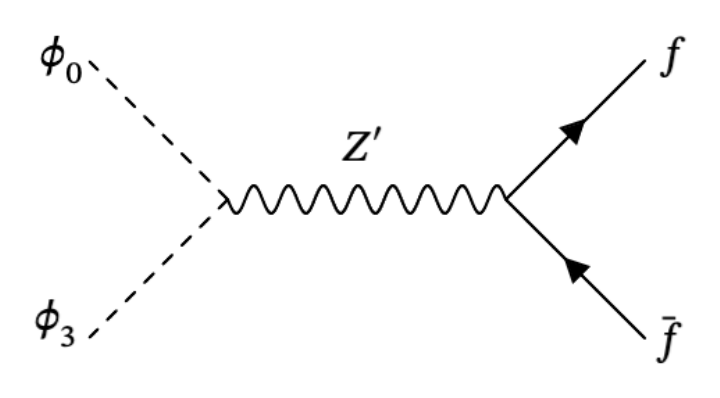}
\includegraphics[scale=0.5]{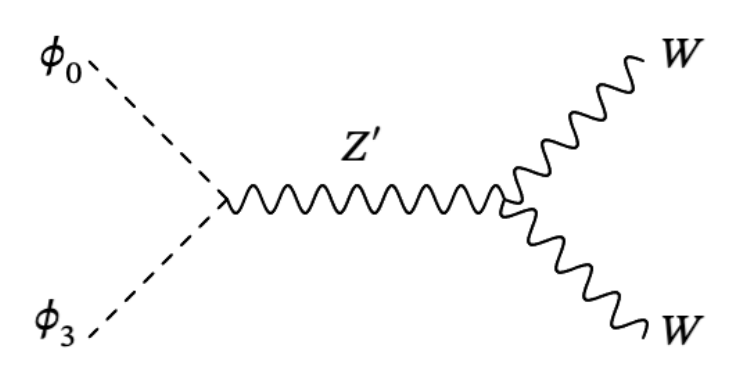}
\includegraphics[scale=0.5]{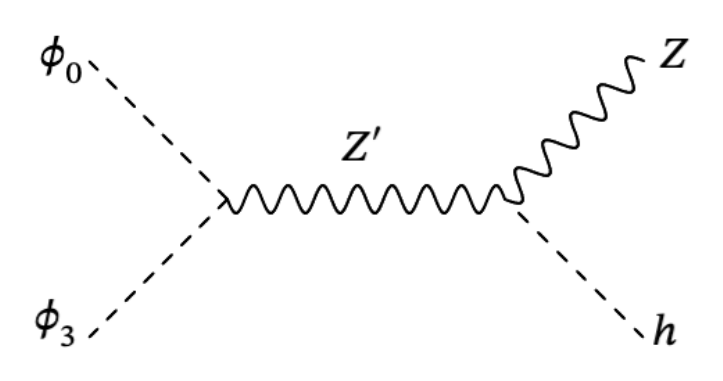}
\caption{Relevant co-annihilation channels for relic abundance calculation.}
\label{fig:phi_03_co-anni}
\end{figure}

The coupled Boltzmann equations for the multicomponent DM are given by the following equations :
\begin{equation}
\frac{dn_0}{dt}+3Hn_0=-\langle\sigma v\rangle_{00}(n_0^2-n_{0_{eq}}^2)-\langle\sigma v\rangle_{03}(n_0n_3-n_{0_{eq}}n_{3_{eq}})
\end{equation}
\begin{equation}
\frac{dn_3}{dt}+3Hn_3=-\langle\sigma v\rangle_{33}(n_3^2-n_{3_{eq}}^2)-\langle\sigma v\rangle_{03}(n_0n_3-n_{0_{eq}}n_{3_{eq}})
\end{equation}
where $n_i$ is the number density of $\phi_i$ and $\langle\sigma v\rangle_{ij}$ is the velocity averaged cross section of annihilation/co-annihilation of $\phi_i$ and $\phi_j$. The expression for the cross sections can be found in many references including \cite{Guo:2010hq,Basak:2013cga,Campbell:2016zbp,Nam:2020byw}. The corresponding Feynman diagrams for $\phi_0$ annihilation are given in figure \ref{fig:phi_0_anni} where $V$ denotes the SM vector bosons. The $\phi_3$ annihilation channels are same as the $\phi_0$ channels except the additional channels of figure \ref{fig:phi_3_anni}. The Feynman diagrams corresponding to the co-annihilation channels are given in figure \ref{fig:phi_03_co-anni}.

\FloatBarrier
These equations have been solved to find the relic \cite{Kolb:1990vq,Srednicki:1988ce,Edsjo:1997bg} by defining an effective velocity averaged cross section in a similar way as in ref.~\cite{Ghosh:2021khk,Bhattacharya:2018cgx} considering appropriate channels of production.

\bibliographystyle{apsrev}
\bibliography{Ref_Paper}

\begin{thebibliography}{74}
\expandafter\ifx\csname natexlab\endcsname\relax\def\natexlab#1{#1}\fi
\expandafter\ifx\csname bibnamefont\endcsname\relax
  \def\bibnamefont#1{#1}\fi
\expandafter\ifx\csname bibfnamefont\endcsname\relax
  \def\bibfnamefont#1{#1}\fi
\expandafter\ifx\csname citenamefont\endcsname\relax
  \def\citenamefont#1{#1}\fi
\expandafter\ifx\csname url\endcsname\relax
  \def\url#1{\texttt{#1}}\fi
\expandafter\ifx\csname urlprefix\endcsname\relax\def\urlprefix{URL }\fi
\providecommand{\bibinfo}[2]{#2}
\providecommand{\eprint}[2][]{\url{#2}}

\bibitem[{\citenamefont{Aghanim et~al.}(2020{\natexlab{a}})}]{Planck:2018vyg}
\bibinfo{author}{\bibfnamefont{N.}~\bibnamefont{Aghanim}} \bibnamefont{et~al.}
  (\bibinfo{collaboration}{Planck}), \bibinfo{journal}{Astron. Astrophys.}
  \textbf{\bibinfo{volume}{641}}, \bibinfo{pages}{A6}
  (\bibinfo{year}{2020}{\natexlab{a}}), \bibinfo{note}{[Erratum:
  Astron.Astrophys. 652, C4 (2021)]}, \eprint{1807.06209}.

\bibitem[{\citenamefont{Hinshaw et~al.}(2013)}]{WMAP:2012nax}
\bibinfo{author}{\bibfnamefont{G.}~\bibnamefont{Hinshaw}} \bibnamefont{et~al.}
  (\bibinfo{collaboration}{WMAP}), \bibinfo{journal}{Astrophys. J. Suppl.}
  \textbf{\bibinfo{volume}{208}}, \bibinfo{pages}{19} (\bibinfo{year}{2013}),
  \eprint{1212.5226}.

\bibitem[{\citenamefont{Bertone et~al.}(2005)\citenamefont{Bertone, Hooper, and
  Silk}}]{Bertone:2004pz}
\bibinfo{author}{\bibfnamefont{G.}~\bibnamefont{Bertone}},
  \bibinfo{author}{\bibfnamefont{D.}~\bibnamefont{Hooper}}, \bibnamefont{and}
  \bibinfo{author}{\bibfnamefont{J.}~\bibnamefont{Silk}},
  \bibinfo{journal}{Phys.Rept.} \textbf{\bibinfo{volume}{405}},
  \bibinfo{pages}{279} (\bibinfo{year}{2005}), \eprint{hep-ph/0404175}.

\bibitem[{\citenamefont{Bergstrom}(2009)}]{Bergstrom:2009ib}
\bibinfo{author}{\bibfnamefont{L.}~\bibnamefont{Bergstrom}},
  \bibinfo{journal}{New J.Phys.} \textbf{\bibinfo{volume}{11}},
  \bibinfo{pages}{105006} (\bibinfo{year}{2009}), \eprint{0903.4849}.

\bibitem[{\citenamefont{Arcadi et~al.}(2018)\citenamefont{Arcadi, Dutra, Ghosh,
  Lindner, Mambrini, Pierre, Profumo, and Queiroz}}]{Arcadi:2017kky}
\bibinfo{author}{\bibfnamefont{G.}~\bibnamefont{Arcadi}},
  \bibinfo{author}{\bibfnamefont{M.}~\bibnamefont{Dutra}},
  \bibinfo{author}{\bibfnamefont{P.}~\bibnamefont{Ghosh}},
  \bibinfo{author}{\bibfnamefont{M.}~\bibnamefont{Lindner}},
  \bibinfo{author}{\bibfnamefont{Y.}~\bibnamefont{Mambrini}},
  \bibinfo{author}{\bibfnamefont{M.}~\bibnamefont{Pierre}},
  \bibinfo{author}{\bibfnamefont{S.}~\bibnamefont{Profumo}}, \bibnamefont{and}
  \bibinfo{author}{\bibfnamefont{F.~S.} \bibnamefont{Queiroz}},
  \bibinfo{journal}{Eur. Phys. J. C} \textbf{\bibinfo{volume}{78}},
  \bibinfo{pages}{203} (\bibinfo{year}{2018}), \eprint{1703.07364}.

\bibitem[{\citenamefont{McDonald}(1994)}]{McDonald:1993ex}
\bibinfo{author}{\bibfnamefont{J.}~\bibnamefont{McDonald}},
  \bibinfo{journal}{Phys.Rev.} \textbf{\bibinfo{volume}{D50}},
  \bibinfo{pages}{3637} (\bibinfo{year}{1994}), \eprint{hep-ph/0702143}.

\bibitem[{\citenamefont{Burgess et~al.}(2001)\citenamefont{Burgess, Pospelov,
  and ter Veldhuis}}]{Burgess:2000yq}
\bibinfo{author}{\bibfnamefont{C.}~\bibnamefont{Burgess}},
  \bibinfo{author}{\bibfnamefont{M.}~\bibnamefont{Pospelov}}, \bibnamefont{and}
  \bibinfo{author}{\bibfnamefont{T.}~\bibnamefont{ter Veldhuis}},
  \bibinfo{journal}{Nucl.Phys.} \textbf{\bibinfo{volume}{B619}},
  \bibinfo{pages}{709} (\bibinfo{year}{2001}), \eprint{hep-ph/0011335}.

\bibitem[{\citenamefont{Guo and Wu}(2010)}]{Guo:2010hq}
\bibinfo{author}{\bibfnamefont{W.-L.} \bibnamefont{Guo}} \bibnamefont{and}
  \bibinfo{author}{\bibfnamefont{Y.-L.} \bibnamefont{Wu}},
  \bibinfo{journal}{JHEP} \textbf{\bibinfo{volume}{1010}}, \bibinfo{pages}{083}
  (\bibinfo{year}{2010}), \eprint{1006.2518}.

\bibitem[{\citenamefont{Bandyopadhyay et~al.}(2010)\citenamefont{Bandyopadhyay,
  Chakraborty, Ghosal, and Majumdar}}]{Bandyopadhyay:2010cc}
\bibinfo{author}{\bibfnamefont{A.}~\bibnamefont{Bandyopadhyay}},
  \bibinfo{author}{\bibfnamefont{S.}~\bibnamefont{Chakraborty}},
  \bibinfo{author}{\bibfnamefont{A.}~\bibnamefont{Ghosal}}, \bibnamefont{and}
  \bibinfo{author}{\bibfnamefont{D.}~\bibnamefont{Majumdar}},
  \bibinfo{journal}{JHEP} \textbf{\bibinfo{volume}{1011}}, \bibinfo{pages}{065}
  (\bibinfo{year}{2010}), \eprint{1003.0809}.

\bibitem[{\citenamefont{He et~al.}(2009)\citenamefont{He, Li, Li, Tandean, and
  Tsai}}]{He:2008qm}
\bibinfo{author}{\bibfnamefont{X.-G.} \bibnamefont{He}},
  \bibinfo{author}{\bibfnamefont{T.}~\bibnamefont{Li}},
  \bibinfo{author}{\bibfnamefont{X.-Q.} \bibnamefont{Li}},
  \bibinfo{author}{\bibfnamefont{J.}~\bibnamefont{Tandean}}, \bibnamefont{and}
  \bibinfo{author}{\bibfnamefont{H.-C.} \bibnamefont{Tsai}},
  \bibinfo{journal}{Phys. Rev. D} \textbf{\bibinfo{volume}{79}},
  \bibinfo{pages}{023521} (\bibinfo{year}{2009}), \eprint{0811.0658}.

\bibitem[{\citenamefont{He et~al.}(2010)\citenamefont{He, Li, Li, Tandean, and
  Tsai}}]{He:2009yd}
\bibinfo{author}{\bibfnamefont{X.-G.} \bibnamefont{He}},
  \bibinfo{author}{\bibfnamefont{T.}~\bibnamefont{Li}},
  \bibinfo{author}{\bibfnamefont{X.-Q.} \bibnamefont{Li}},
  \bibinfo{author}{\bibfnamefont{J.}~\bibnamefont{Tandean}}, \bibnamefont{and}
  \bibinfo{author}{\bibfnamefont{H.-C.} \bibnamefont{Tsai}},
  \bibinfo{journal}{Phys. Lett. B} \textbf{\bibinfo{volume}{688}},
  \bibinfo{pages}{332} (\bibinfo{year}{2010}), \eprint{0912.4722}.

\bibitem[{\citenamefont{Cline et~al.}(2013)\citenamefont{Cline, Kainulainen,
  Scott, and Weniger}}]{Cline:2013gha}
\bibinfo{author}{\bibfnamefont{J.~M.} \bibnamefont{Cline}},
  \bibinfo{author}{\bibfnamefont{K.}~\bibnamefont{Kainulainen}},
  \bibinfo{author}{\bibfnamefont{P.}~\bibnamefont{Scott}}, \bibnamefont{and}
  \bibinfo{author}{\bibfnamefont{C.}~\bibnamefont{Weniger}},
  \bibinfo{journal}{Phys. Rev. D} \textbf{\bibinfo{volume}{88}},
  \bibinfo{pages}{055025} (\bibinfo{year}{2013}), \bibinfo{note}{[Erratum:
  Phys.Rev.D 92, 039906 (2015)]}, \eprint{1306.4710}.

\bibitem[{\citenamefont{Abada et~al.}(2011)\citenamefont{Abada, Ghaffor, and
  Nasri}}]{Abada:2011qb}
\bibinfo{author}{\bibfnamefont{A.}~\bibnamefont{Abada}},
  \bibinfo{author}{\bibfnamefont{D.}~\bibnamefont{Ghaffor}}, \bibnamefont{and}
  \bibinfo{author}{\bibfnamefont{S.}~\bibnamefont{Nasri}},
  \bibinfo{journal}{Phys. Rev. D} \textbf{\bibinfo{volume}{83}},
  \bibinfo{pages}{095021} (\bibinfo{year}{2011}), \eprint{1101.0365}.

\bibitem[{\citenamefont{Abada and Nasri}(2012)}]{Abada:2012hf}
\bibinfo{author}{\bibfnamefont{A.}~\bibnamefont{Abada}} \bibnamefont{and}
  \bibinfo{author}{\bibfnamefont{S.}~\bibnamefont{Nasri}},
  \bibinfo{journal}{Phys. Rev. D} \textbf{\bibinfo{volume}{85}},
  \bibinfo{pages}{075009} (\bibinfo{year}{2012}), \eprint{1201.1413}.

\bibitem[{\citenamefont{Arhrib and Maniatis}(2019)}]{Arhrib:2018eex}
\bibinfo{author}{\bibfnamefont{A.}~\bibnamefont{Arhrib}} \bibnamefont{and}
  \bibinfo{author}{\bibfnamefont{M.}~\bibnamefont{Maniatis}},
  \bibinfo{journal}{Phys. Lett. B} \textbf{\bibinfo{volume}{796}},
  \bibinfo{pages}{15} (\bibinfo{year}{2019}), \eprint{1807.03554}.

\bibitem[{\citenamefont{Hamada et~al.}(2021)\citenamefont{Hamada, Kawai, Oda,
  and Yagyu}}]{Hamada:2020wjh}
\bibinfo{author}{\bibfnamefont{Y.}~\bibnamefont{Hamada}},
  \bibinfo{author}{\bibfnamefont{H.}~\bibnamefont{Kawai}},
  \bibinfo{author}{\bibfnamefont{K.-y.} \bibnamefont{Oda}}, \bibnamefont{and}
  \bibinfo{author}{\bibfnamefont{K.}~\bibnamefont{Yagyu}},
  \bibinfo{journal}{JHEP} \textbf{\bibinfo{volume}{01}}, \bibinfo{pages}{087}
  (\bibinfo{year}{2021}), \eprint{2008.08700}.

\bibitem[{\citenamefont{Modak et~al.}(2015)\citenamefont{Modak, Majumdar, and
  Rakshit}}]{Modak:2013jya}
\bibinfo{author}{\bibfnamefont{K.~P.} \bibnamefont{Modak}},
  \bibinfo{author}{\bibfnamefont{D.}~\bibnamefont{Majumdar}}, \bibnamefont{and}
  \bibinfo{author}{\bibfnamefont{S.}~\bibnamefont{Rakshit}},
  \bibinfo{journal}{JCAP} \textbf{\bibinfo{volume}{03}}, \bibinfo{pages}{011}
  (\bibinfo{year}{2015}), \eprint{1312.7488}.

\bibitem[{\citenamefont{Maniatis}(2021)}]{Maniatis:2020ois}
\bibinfo{author}{\bibfnamefont{M.}~\bibnamefont{Maniatis}},
  \bibinfo{journal}{Phys. Rev. D} \textbf{\bibinfo{volume}{103}},
  \bibinfo{pages}{015010} (\bibinfo{year}{2021}), \eprint{2005.13443}.

\bibitem[{\citenamefont{Bhattacharya
  et~al.}(2017{\natexlab{a}})\citenamefont{Bhattacharya, Ghosh, Maity, and
  Ray}}]{Bhattacharya:2017fid}
\bibinfo{author}{\bibfnamefont{S.}~\bibnamefont{Bhattacharya}},
  \bibinfo{author}{\bibfnamefont{P.}~\bibnamefont{Ghosh}},
  \bibinfo{author}{\bibfnamefont{T.~N.} \bibnamefont{Maity}}, \bibnamefont{and}
  \bibinfo{author}{\bibfnamefont{T.~S.} \bibnamefont{Ray}},
  \bibinfo{journal}{JHEP} \textbf{\bibinfo{volume}{10}}, \bibinfo{pages}{088}
  (\bibinfo{year}{2017}{\natexlab{a}}), \eprint{1706.04699}.

\bibitem[{\citenamefont{Maity and Ray}(2020)}]{Maity:2019hre}
\bibinfo{author}{\bibfnamefont{T.~N.} \bibnamefont{Maity}} \bibnamefont{and}
  \bibinfo{author}{\bibfnamefont{T.~S.} \bibnamefont{Ray}},
  \bibinfo{journal}{Phys. Rev. D} \textbf{\bibinfo{volume}{101}},
  \bibinfo{pages}{103013} (\bibinfo{year}{2020}), \eprint{1908.10343}.

\bibitem[{\citenamefont{D\'\i{}az~S\'aez
  et~al.}(2021)\citenamefont{D\'\i{}az~S\'aez, M\"ohling, and
  St\"ockinger}}]{DiazSaez:2021pfw}
\bibinfo{author}{\bibfnamefont{B.}~\bibnamefont{D\'\i{}az~S\'aez}},
  \bibinfo{author}{\bibfnamefont{K.}~\bibnamefont{M\"ohling}},
  \bibnamefont{and}
  \bibinfo{author}{\bibfnamefont{D.}~\bibnamefont{St\"ockinger}},
  \bibinfo{journal}{JCAP} \textbf{\bibinfo{volume}{10}}, \bibinfo{pages}{027}
  (\bibinfo{year}{2021}), \eprint{2103.17064}.

\bibitem[{\citenamefont{Basak et~al.}(2021)\citenamefont{Basak, Coleppa, and
  Loho}}]{Basak:2021tnj}
\bibinfo{author}{\bibfnamefont{T.}~\bibnamefont{Basak}},
  \bibinfo{author}{\bibfnamefont{B.}~\bibnamefont{Coleppa}}, \bibnamefont{and}
  \bibinfo{author}{\bibfnamefont{K.}~\bibnamefont{Loho}},
  \bibinfo{journal}{JHEP} \textbf{\bibinfo{volume}{06}}, \bibinfo{pages}{104}
  (\bibinfo{year}{2021}), \eprint{2105.09044}.

\bibitem[{\citenamefont{Lopez~Honorez et~al.}(2007)\citenamefont{Lopez~Honorez,
  Nezri, Oliver, and Tytgat}}]{LopezHonorez:2006gr}
\bibinfo{author}{\bibfnamefont{L.}~\bibnamefont{Lopez~Honorez}},
  \bibinfo{author}{\bibfnamefont{E.}~\bibnamefont{Nezri}},
  \bibinfo{author}{\bibfnamefont{J.~F.} \bibnamefont{Oliver}},
  \bibnamefont{and} \bibinfo{author}{\bibfnamefont{M.~H.~G.}
  \bibnamefont{Tytgat}}, \bibinfo{journal}{JCAP} \textbf{\bibinfo{volume}{02}},
  \bibinfo{pages}{028} (\bibinfo{year}{2007}), \eprint{hep-ph/0612275}.

\bibitem[{\citenamefont{Belyaev et~al.}(2018)\citenamefont{Belyaev,
  Cacciapaglia, Ivanov, Rojas-Abatte, and Thomas}}]{Belyaev:2016lok}
\bibinfo{author}{\bibfnamefont{A.}~\bibnamefont{Belyaev}},
  \bibinfo{author}{\bibfnamefont{G.}~\bibnamefont{Cacciapaglia}},
  \bibinfo{author}{\bibfnamefont{I.~P.} \bibnamefont{Ivanov}},
  \bibinfo{author}{\bibfnamefont{F.}~\bibnamefont{Rojas-Abatte}},
  \bibnamefont{and} \bibinfo{author}{\bibfnamefont{M.}~\bibnamefont{Thomas}},
  \bibinfo{journal}{Phys. Rev. D} \textbf{\bibinfo{volume}{97}},
  \bibinfo{pages}{035011} (\bibinfo{year}{2018}), \eprint{1612.00511}.

\bibitem[{\citenamefont{Aoki et~al.}(2010)\citenamefont{Aoki, Kanemura, and
  Seto}}]{Aoki:2009pf}
\bibinfo{author}{\bibfnamefont{M.}~\bibnamefont{Aoki}},
  \bibinfo{author}{\bibfnamefont{S.}~\bibnamefont{Kanemura}}, \bibnamefont{and}
  \bibinfo{author}{\bibfnamefont{O.}~\bibnamefont{Seto}},
  \bibinfo{journal}{Phys. Lett. B} \textbf{\bibinfo{volume}{685}},
  \bibinfo{pages}{313} (\bibinfo{year}{2010}), \eprint{0912.5536}.

\bibitem[{\citenamefont{Bhattacharya
  et~al.}(2020{\natexlab{a}})\citenamefont{Bhattacharya, Ghosh, Saha, and
  Sil}}]{Bhattacharya:2019fgs}
\bibinfo{author}{\bibfnamefont{S.}~\bibnamefont{Bhattacharya}},
  \bibinfo{author}{\bibfnamefont{P.}~\bibnamefont{Ghosh}},
  \bibinfo{author}{\bibfnamefont{A.~K.} \bibnamefont{Saha}}, \bibnamefont{and}
  \bibinfo{author}{\bibfnamefont{A.}~\bibnamefont{Sil}},
  \bibinfo{journal}{JHEP} \textbf{\bibinfo{volume}{03}}, \bibinfo{pages}{090}
  (\bibinfo{year}{2020}{\natexlab{a}}), \eprint{1905.12583}.

\bibitem[{\citenamefont{Rojas-Abatte et~al.}(2017)\citenamefont{Rojas-Abatte,
  Mora, Urbina, and Zerwekh}}]{Rojas-Abatte:2017hqm}
\bibinfo{author}{\bibfnamefont{F.}~\bibnamefont{Rojas-Abatte}},
  \bibinfo{author}{\bibfnamefont{M.~L.} \bibnamefont{Mora}},
  \bibinfo{author}{\bibfnamefont{J.}~\bibnamefont{Urbina}}, \bibnamefont{and}
  \bibinfo{author}{\bibfnamefont{A.~R.} \bibnamefont{Zerwekh}},
  \bibinfo{journal}{Phys. Rev. D} \textbf{\bibinfo{volume}{96}},
  \bibinfo{pages}{095025} (\bibinfo{year}{2017}), \eprint{1707.04543}.

\bibitem[{\citenamefont{Bhat and Adhikari}(2020)}]{Bhat:2019yqo}
\bibinfo{author}{\bibfnamefont{I.~A.} \bibnamefont{Bhat}} \bibnamefont{and}
  \bibinfo{author}{\bibfnamefont{R.}~\bibnamefont{Adhikari}},
  \bibinfo{journal}{Phys. Rev. D} \textbf{\bibinfo{volume}{101}},
  \bibinfo{pages}{075030} (\bibinfo{year}{2020}), \eprint{1906.10185}.

\bibitem[{\citenamefont{Nam}(2020)}]{Nam:2020byw}
\bibinfo{author}{\bibfnamefont{C.~H.} \bibnamefont{Nam}},
  \bibinfo{journal}{Eur. Phys. J. C} \textbf{\bibinfo{volume}{80}},
  \bibinfo{pages}{1114} (\bibinfo{year}{2020}), \eprint{2011.11207}.

\bibitem[{\citenamefont{Okada and Seto}(2010)}]{Okada:2010wd}
\bibinfo{author}{\bibfnamefont{N.}~\bibnamefont{Okada}} \bibnamefont{and}
  \bibinfo{author}{\bibfnamefont{O.}~\bibnamefont{Seto}},
  \bibinfo{journal}{Phys. Rev. D} \textbf{\bibinfo{volume}{82}},
  \bibinfo{pages}{023507} (\bibinfo{year}{2010}), \eprint{1002.2525}.

\bibitem[{\citenamefont{Basak and Mondal}(2014)}]{Basak:2013cga}
\bibinfo{author}{\bibfnamefont{T.}~\bibnamefont{Basak}} \bibnamefont{and}
  \bibinfo{author}{\bibfnamefont{T.}~\bibnamefont{Mondal}},
  \bibinfo{journal}{Phys. Rev. D} \textbf{\bibinfo{volume}{89}},
  \bibinfo{pages}{063527} (\bibinfo{year}{2014}), \eprint{1308.0023}.

\bibitem[{\citenamefont{Okada and Seto}(2020)}]{Okada:2019sbb}
\bibinfo{author}{\bibfnamefont{N.}~\bibnamefont{Okada}} \bibnamefont{and}
  \bibinfo{author}{\bibfnamefont{O.}~\bibnamefont{Seto}},
  \bibinfo{journal}{Phys. Rev. D} \textbf{\bibinfo{volume}{101}},
  \bibinfo{pages}{023522} (\bibinfo{year}{2020}), \eprint{1908.09277}.

\bibitem[{\citenamefont{Biswas et~al.}(2016)\citenamefont{Biswas, Choubey, and
  Khan}}]{Biswas:2016yan}
\bibinfo{author}{\bibfnamefont{A.}~\bibnamefont{Biswas}},
  \bibinfo{author}{\bibfnamefont{S.}~\bibnamefont{Choubey}}, \bibnamefont{and}
  \bibinfo{author}{\bibfnamefont{S.}~\bibnamefont{Khan}},
  \bibinfo{journal}{JHEP} \textbf{\bibinfo{volume}{09}}, \bibinfo{pages}{147}
  (\bibinfo{year}{2016}), \eprint{1608.04194}.

\bibitem[{\citenamefont{Bandyopadhyay
  et~al.}(2018{\natexlab{a}})\citenamefont{Bandyopadhyay, Chun, and
  Mandal}}]{Bandyopadhyay:2017bgh}
\bibinfo{author}{\bibfnamefont{P.}~\bibnamefont{Bandyopadhyay}},
  \bibinfo{author}{\bibfnamefont{E.~J.} \bibnamefont{Chun}}, \bibnamefont{and}
  \bibinfo{author}{\bibfnamefont{R.}~\bibnamefont{Mandal}},
  \bibinfo{journal}{Phys. Rev. D} \textbf{\bibinfo{volume}{97}},
  \bibinfo{pages}{015001} (\bibinfo{year}{2018}{\natexlab{a}}),
  \eprint{1707.00874}.

\bibitem[{\citenamefont{Bhattacharyya and Das}(2016)}]{Bhattacharyya:2015nca}
\bibinfo{author}{\bibfnamefont{G.}~\bibnamefont{Bhattacharyya}}
  \bibnamefont{and} \bibinfo{author}{\bibfnamefont{D.}~\bibnamefont{Das}},
  \bibinfo{journal}{Pramana} \textbf{\bibinfo{volume}{87}}, \bibinfo{pages}{40}
  (\bibinfo{year}{2016}), \eprint{1507.06424}.

\bibitem[{ATL(2020)}]{ATLAS-CONF-2020-052}
\bibinfo{type}{Tech. Rep.} \bibinfo{number}{ATLAS-CONF-2020-052},
  \bibinfo{institution}{CERN}, \bibinfo{address}{Geneva}
  (\bibinfo{year}{2020}), \urlprefix\url{http://cds.cern.ch/record/2743055}.

\bibitem[{\citenamefont{Sirunyan et~al.}(2019)}]{CMS:2018yfx}
\bibinfo{author}{\bibfnamefont{A.~M.} \bibnamefont{Sirunyan}}
  \bibnamefont{et~al.} (\bibinfo{collaboration}{CMS}), \bibinfo{journal}{Phys.
  Lett. B} \textbf{\bibinfo{volume}{793}}, \bibinfo{pages}{520}
  (\bibinfo{year}{2019}), \eprint{1809.05937}.

\bibitem[{\citenamefont{Aad et~al.}(2023{\natexlab{a}})}]{ATLAS:2022tnm}
\bibinfo{author}{\bibfnamefont{G.}~\bibnamefont{Aad}} \bibnamefont{et~al.}
  (\bibinfo{collaboration}{ATLAS}), \bibinfo{journal}{JHEP}
  \textbf{\bibinfo{volume}{07}}, \bibinfo{pages}{088}
  (\bibinfo{year}{2023}{\natexlab{a}}), \eprint{2207.00348}.

\bibitem[{CMS(2020)}]{CMS-PAS-HIG-19-015}
\bibinfo{type}{Tech. Rep.}, \bibinfo{institution}{CERN},
  \bibinfo{address}{Geneva} (\bibinfo{year}{2020}),
  \urlprefix\url{https://cds.cern.ch/record/2725142}.

\bibitem[{\citenamefont{Electroweak}(2003)}]{Electroweak:2003ram}
\bibinfo{author}{\bibfnamefont{t.~S.} \bibnamefont{Electroweak}}
  (\bibinfo{collaboration}{LEP, ALEPH, DELPHI, L3, OPAL, LEP Electroweak
  Working Group, SLD Electroweak Group, SLD Heavy Flavor Group})
  (\bibinfo{year}{2003}), \eprint{hep-ex/0312023}.

\bibitem[{\citenamefont{Abe et~al.}(1995)}]{CDF:1994kud}
\bibinfo{author}{\bibfnamefont{F.}~\bibnamefont{Abe}} \bibnamefont{et~al.}
  (\bibinfo{collaboration}{CDF}), \bibinfo{journal}{Phys. Rev. D}
  \textbf{\bibinfo{volume}{51}}, \bibinfo{pages}{R949} (\bibinfo{year}{1995}).

\bibitem[{\citenamefont{Abe et~al.}(1997)}]{CDF:1997wdd}
\bibinfo{author}{\bibfnamefont{F.}~\bibnamefont{Abe}} \bibnamefont{et~al.}
  (\bibinfo{collaboration}{CDF}), \bibinfo{journal}{Phys. Rev. Lett.}
  \textbf{\bibinfo{volume}{79}}, \bibinfo{pages}{2192} (\bibinfo{year}{1997}).

\bibitem[{\citenamefont{Abachi et~al.}(1996)}]{D0:1996qsx}
\bibinfo{author}{\bibfnamefont{S.}~\bibnamefont{Abachi}} \bibnamefont{et~al.}
  (\bibinfo{collaboration}{D0}), \bibinfo{journal}{Phys. Lett. B}
  \textbf{\bibinfo{volume}{385}}, \bibinfo{pages}{471} (\bibinfo{year}{1996}).

\bibitem[{\citenamefont{Abazov et~al.}(2001)}]{D0:2001zjx}
\bibinfo{author}{\bibfnamefont{V.~M.} \bibnamefont{Abazov}}
  \bibnamefont{et~al.} (\bibinfo{collaboration}{D0}), \bibinfo{journal}{Phys.
  Rev. Lett.} \textbf{\bibinfo{volume}{87}}, \bibinfo{pages}{061802}
  (\bibinfo{year}{2001}), \eprint{hep-ex/0102048}.

\bibitem[{\citenamefont{Carena et~al.}(2004)\citenamefont{Carena, Daleo,
  Dobrescu, and Tait}}]{Carena:2004xs}
\bibinfo{author}{\bibfnamefont{M.}~\bibnamefont{Carena}},
  \bibinfo{author}{\bibfnamefont{A.}~\bibnamefont{Daleo}},
  \bibinfo{author}{\bibfnamefont{B.~A.} \bibnamefont{Dobrescu}},
  \bibnamefont{and} \bibinfo{author}{\bibfnamefont{T.~M.~P.}
  \bibnamefont{Tait}}, \bibinfo{journal}{Phys. Rev. D}
  \textbf{\bibinfo{volume}{70}}, \bibinfo{pages}{093009}
  (\bibinfo{year}{2004}), \eprint{hep-ph/0408098}.

\bibitem[{\citenamefont{Bandyopadhyay
  et~al.}(2018{\natexlab{b}})\citenamefont{Bandyopadhyay, Bhattacharyya, Das,
  and Raychaudhuri}}]{Bandyopadhyay:2018cwu}
\bibinfo{author}{\bibfnamefont{T.}~\bibnamefont{Bandyopadhyay}},
  \bibinfo{author}{\bibfnamefont{G.}~\bibnamefont{Bhattacharyya}},
  \bibinfo{author}{\bibfnamefont{D.}~\bibnamefont{Das}}, \bibnamefont{and}
  \bibinfo{author}{\bibfnamefont{A.}~\bibnamefont{Raychaudhuri}},
  \bibinfo{journal}{Phys. Rev. D} \textbf{\bibinfo{volume}{98}},
  \bibinfo{pages}{035027} (\bibinfo{year}{2018}{\natexlab{b}}),
  \eprint{1803.07989}.

\bibitem[{\citenamefont{Pappadopulo et~al.}(2014)\citenamefont{Pappadopulo,
  Thamm, Torre, and Wulzer}}]{Pappadopulo:2014qza}
\bibinfo{author}{\bibfnamefont{D.}~\bibnamefont{Pappadopulo}},
  \bibinfo{author}{\bibfnamefont{A.}~\bibnamefont{Thamm}},
  \bibinfo{author}{\bibfnamefont{R.}~\bibnamefont{Torre}}, \bibnamefont{and}
  \bibinfo{author}{\bibfnamefont{A.}~\bibnamefont{Wulzer}},
  \bibinfo{journal}{JHEP} \textbf{\bibinfo{volume}{09}}, \bibinfo{pages}{060}
  (\bibinfo{year}{2014}), \eprint{1402.4431}.

\bibitem[{\citenamefont{Fl\'orez et~al.}(2017)\citenamefont{Fl\'orez, Gurrola,
  Johns, Oh, Sheldon, Teague, and Weiler}}]{Florez:2016uob}
\bibinfo{author}{\bibfnamefont{A.}~\bibnamefont{Fl\'orez}},
  \bibinfo{author}{\bibfnamefont{A.}~\bibnamefont{Gurrola}},
  \bibinfo{author}{\bibfnamefont{W.}~\bibnamefont{Johns}},
  \bibinfo{author}{\bibfnamefont{Y.~D.} \bibnamefont{Oh}},
  \bibinfo{author}{\bibfnamefont{P.}~\bibnamefont{Sheldon}},
  \bibinfo{author}{\bibfnamefont{D.}~\bibnamefont{Teague}}, \bibnamefont{and}
  \bibinfo{author}{\bibfnamefont{T.}~\bibnamefont{Weiler}},
  \bibinfo{journal}{Phys. Lett. B} \textbf{\bibinfo{volume}{767}},
  \bibinfo{pages}{126} (\bibinfo{year}{2017}), \eprint{1609.09765}.

\bibitem[{\citenamefont{Aad et~al.}(2023{\natexlab{b}})}]{ATLAS:2022enb}
\bibinfo{author}{\bibfnamefont{G.}~\bibnamefont{Aad}} \bibnamefont{et~al.}
  (\bibinfo{collaboration}{ATLAS}), \bibinfo{journal}{JHEP}
  \textbf{\bibinfo{volume}{06}}, \bibinfo{pages}{016}
  (\bibinfo{year}{2023}{\natexlab{b}}), \eprint{2207.00230}.

\bibitem[{\citenamefont{Sirunyan et~al.}(2021)}]{CMS:2021fyk}
\bibinfo{author}{\bibfnamefont{A.~M.} \bibnamefont{Sirunyan}}
  \bibnamefont{et~al.} (\bibinfo{collaboration}{CMS}), \bibinfo{journal}{Eur.
  Phys. J. C} \textbf{\bibinfo{volume}{81}}, \bibinfo{pages}{688}
  (\bibinfo{year}{2021}), \eprint{2102.08198}.

\bibitem[{\citenamefont{Aaboud et~al.}(2018)}]{ATLAS:2017jag}
\bibinfo{author}{\bibfnamefont{M.}~\bibnamefont{Aaboud}} \bibnamefont{et~al.}
  (\bibinfo{collaboration}{ATLAS}), \bibinfo{journal}{JHEP}
  \textbf{\bibinfo{volume}{03}}, \bibinfo{pages}{042} (\bibinfo{year}{2018}),
  \eprint{1710.07235}.

\bibitem[{\citenamefont{Tumasyan et~al.}(2022)}]{CMS:2021klu}
\bibinfo{author}{\bibfnamefont{A.}~\bibnamefont{Tumasyan}} \bibnamefont{et~al.}
  (\bibinfo{collaboration}{CMS}), \bibinfo{journal}{Phys. Rev. D}
  \textbf{\bibinfo{volume}{105}}, \bibinfo{pages}{032008}
  (\bibinfo{year}{2022}), \eprint{2109.06055}.

\bibitem[{\citenamefont{Aaboud et~al.}(2017)}]{ATLAS:2017fih}
\bibinfo{author}{\bibfnamefont{M.}~\bibnamefont{Aaboud}} \bibnamefont{et~al.}
  (\bibinfo{collaboration}{ATLAS}), \bibinfo{journal}{JHEP}
  \textbf{\bibinfo{volume}{10}}, \bibinfo{pages}{182} (\bibinfo{year}{2017}),
  \eprint{1707.02424}.

\bibitem[{\citenamefont{Sirunyan et~al.}(2018)}]{CMS:2018ipm}
\bibinfo{author}{\bibfnamefont{A.~M.} \bibnamefont{Sirunyan}}
  \bibnamefont{et~al.} (\bibinfo{collaboration}{CMS}), \bibinfo{journal}{JHEP}
  \textbf{\bibinfo{volume}{06}}, \bibinfo{pages}{120} (\bibinfo{year}{2018}),
  \eprint{1803.06292}.

\bibitem[{\citenamefont{Cirelli et~al.}(2006)\citenamefont{Cirelli, Fornengo,
  and Strumia}}]{Cirelli:2005uq}
\bibinfo{author}{\bibfnamefont{M.}~\bibnamefont{Cirelli}},
  \bibinfo{author}{\bibfnamefont{N.}~\bibnamefont{Fornengo}}, \bibnamefont{and}
  \bibinfo{author}{\bibfnamefont{A.}~\bibnamefont{Strumia}},
  \bibinfo{journal}{Nucl. Phys. B} \textbf{\bibinfo{volume}{753}},
  \bibinfo{pages}{178} (\bibinfo{year}{2006}), \eprint{hep-ph/0512090}.

\bibitem[{\citenamefont{Aghanim et~al.}(2020{\natexlab{b}})}]{planck2018}
\bibinfo{author}{\bibfnamefont{N.}~\bibnamefont{Aghanim}} \bibnamefont{et~al.}
  (\bibinfo{collaboration}{Planck}), \bibinfo{journal}{Astron. Astrophys.}
  \textbf{\bibinfo{volume}{641}}, \bibinfo{pages}{A6}
  (\bibinfo{year}{2020}{\natexlab{b}}), \eprint{1807.06209}.

\bibitem[{\citenamefont{Aprile et~al.}(2018)}]{Aprile:2018dbl}
\bibinfo{author}{\bibfnamefont{E.}~\bibnamefont{Aprile}} \bibnamefont{et~al.}
  (\bibinfo{collaboration}{XENON}), \bibinfo{journal}{Phys. Rev. Lett.}
  \textbf{\bibinfo{volume}{121}}, \bibinfo{pages}{111302}
  (\bibinfo{year}{2018}), \eprint{1805.12562}.

\bibitem[{\citenamefont{Aprile et~al.}(2022)}]{XENON:2022avm}
\bibinfo{author}{\bibfnamefont{E.}~\bibnamefont{Aprile}} \bibnamefont{et~al.}
  (\bibinfo{collaboration}{XENON}) (\bibinfo{year}{2022}), \eprint{2210.07591}.

\bibitem[{\citenamefont{Meng et~al.}(2021)}]{PandaX-4T:2021bab}
\bibinfo{author}{\bibfnamefont{Y.}~\bibnamefont{Meng}} \bibnamefont{et~al.}
  (\bibinfo{collaboration}{PandaX-4T}), \bibinfo{journal}{Phys. Rev. Lett.}
  \textbf{\bibinfo{volume}{127}}, \bibinfo{pages}{261802}
  (\bibinfo{year}{2021}), \eprint{2107.13438}.

\bibitem[{\citenamefont{Aalbers et~al.}(2022)}]{LZ:2022ufs}
\bibinfo{author}{\bibfnamefont{J.}~\bibnamefont{Aalbers}} \bibnamefont{et~al.}
  (\bibinfo{collaboration}{LZ}) (\bibinfo{year}{2022}), \eprint{2207.03764}.

\bibitem[{\citenamefont{Ellis et~al.}(2000)\citenamefont{Ellis, Ferstl, and
  Olive}}]{Ellis:2000ds}
\bibinfo{author}{\bibfnamefont{J.~R.} \bibnamefont{Ellis}},
  \bibinfo{author}{\bibfnamefont{A.}~\bibnamefont{Ferstl}}, \bibnamefont{and}
  \bibinfo{author}{\bibfnamefont{K.~A.} \bibnamefont{Olive}},
  \bibinfo{journal}{Phys.Lett.} \textbf{\bibinfo{volume}{B481}},
  \bibinfo{pages}{304} (\bibinfo{year}{2000}), \eprint{hep-ph/0001005}.

\bibitem[{\citenamefont{Profumo et~al.}(2009)\citenamefont{Profumo, Sigurdson,
  and Ubaldi}}]{Profumo:2009tb}
\bibinfo{author}{\bibfnamefont{S.}~\bibnamefont{Profumo}},
  \bibinfo{author}{\bibfnamefont{K.}~\bibnamefont{Sigurdson}},
  \bibnamefont{and} \bibinfo{author}{\bibfnamefont{L.}~\bibnamefont{Ubaldi}},
  \bibinfo{journal}{JCAP} \textbf{\bibinfo{volume}{12}}, \bibinfo{pages}{016}
  (\bibinfo{year}{2009}), \eprint{0907.4374}.

\bibitem[{\citenamefont{Bhattacharya
  et~al.}(2017{\natexlab{b}})\citenamefont{Bhattacharya, Poulose, and
  Ghosh}}]{Bhattacharya:2016ysw}
\bibinfo{author}{\bibfnamefont{S.}~\bibnamefont{Bhattacharya}},
  \bibinfo{author}{\bibfnamefont{P.}~\bibnamefont{Poulose}}, \bibnamefont{and}
  \bibinfo{author}{\bibfnamefont{P.}~\bibnamefont{Ghosh}},
  \bibinfo{journal}{JCAP} \textbf{\bibinfo{volume}{04}}, \bibinfo{pages}{043}
  (\bibinfo{year}{2017}{\natexlab{b}}), \eprint{1607.08461}.

\bibitem[{\citenamefont{Herrero-Garcia
  et~al.}(2017)\citenamefont{Herrero-Garcia, Scaffidi, White, and
  Williams}}]{Herrero-Garcia:2017vrl}
\bibinfo{author}{\bibfnamefont{J.}~\bibnamefont{Herrero-Garcia}},
  \bibinfo{author}{\bibfnamefont{A.}~\bibnamefont{Scaffidi}},
  \bibinfo{author}{\bibfnamefont{M.}~\bibnamefont{White}}, \bibnamefont{and}
  \bibinfo{author}{\bibfnamefont{A.~G.} \bibnamefont{Williams}},
  \bibinfo{journal}{JCAP} \textbf{\bibinfo{volume}{11}}, \bibinfo{pages}{021}
  (\bibinfo{year}{2017}), \eprint{1709.01945}.

\bibitem[{\citenamefont{Bhattacharya
  et~al.}(2020{\natexlab{b}})\citenamefont{Bhattacharya, Chakrabarty, Roshan,
  and Sil}}]{Bhattacharya:2019tqq}
\bibinfo{author}{\bibfnamefont{S.}~\bibnamefont{Bhattacharya}},
  \bibinfo{author}{\bibfnamefont{N.}~\bibnamefont{Chakrabarty}},
  \bibinfo{author}{\bibfnamefont{R.}~\bibnamefont{Roshan}}, \bibnamefont{and}
  \bibinfo{author}{\bibfnamefont{A.}~\bibnamefont{Sil}},
  \bibinfo{journal}{JCAP} \textbf{\bibinfo{volume}{04}}, \bibinfo{pages}{013}
  (\bibinfo{year}{2020}{\natexlab{b}}), \eprint{1910.00612}.

\bibitem[{\citenamefont{Billard et~al.}(2022)}]{Billard:2021uyg}
\bibinfo{author}{\bibfnamefont{J.}~\bibnamefont{Billard}} \bibnamefont{et~al.},
  \bibinfo{journal}{Rept. Prog. Phys.} \textbf{\bibinfo{volume}{85}},
  \bibinfo{pages}{056201} (\bibinfo{year}{2022}), \eprint{2104.07634}.

\bibitem[{\citenamefont{O'Hare}(2021)}]{OHare:2021utq}
\bibinfo{author}{\bibfnamefont{C.~A.~J.} \bibnamefont{O'Hare}},
  \bibinfo{journal}{Phys. Rev. Lett.} \textbf{\bibinfo{volume}{127}},
  \bibinfo{pages}{251802} (\bibinfo{year}{2021}), \eprint{2109.03116}.

\bibitem[{\citenamefont{Gunion et~al.}(2000)\citenamefont{Gunion, Haber, Kane,
  and Dawson}}]{Gunion:1989we}
\bibinfo{author}{\bibfnamefont{J.~F.} \bibnamefont{Gunion}},
  \bibinfo{author}{\bibfnamefont{H.~E.} \bibnamefont{Haber}},
  \bibinfo{author}{\bibfnamefont{G.~L.} \bibnamefont{Kane}}, \bibnamefont{and}
  \bibinfo{author}{\bibfnamefont{S.}~\bibnamefont{Dawson}},
  \bibinfo{journal}{Front.Phys.} \textbf{\bibinfo{volume}{80}},
  \bibinfo{pages}{1} (\bibinfo{year}{2000}).

\bibitem[{\citenamefont{Campbell et~al.}(2017)\citenamefont{Campbell, Godfrey,
  Logan, and Poulin}}]{Campbell:2016zbp}
\bibinfo{author}{\bibfnamefont{R.}~\bibnamefont{Campbell}},
  \bibinfo{author}{\bibfnamefont{S.}~\bibnamefont{Godfrey}},
  \bibinfo{author}{\bibfnamefont{H.~E.} \bibnamefont{Logan}}, \bibnamefont{and}
  \bibinfo{author}{\bibfnamefont{A.}~\bibnamefont{Poulin}},
  \bibinfo{journal}{Phys. Rev. D} \textbf{\bibinfo{volume}{95}},
  \bibinfo{pages}{016005} (\bibinfo{year}{2017}), \eprint{1610.08097}.

\bibitem[{\citenamefont{Kolb and Turner}(1990)}]{Kolb:1990vq}
\bibinfo{author}{\bibfnamefont{E.~W.} \bibnamefont{Kolb}} \bibnamefont{and}
  \bibinfo{author}{\bibfnamefont{M.~S.} \bibnamefont{Turner}},
  \emph{\bibinfo{title}{{The Early Universe}}}, vol.~\bibinfo{volume}{69}
  (\bibinfo{year}{1990}), ISBN \bibinfo{isbn}{978-0-201-62674-2}.

\bibitem[{\citenamefont{Srednicki et~al.}(1988)\citenamefont{Srednicki,
  Watkins, and Olive}}]{Srednicki:1988ce}
\bibinfo{author}{\bibfnamefont{M.}~\bibnamefont{Srednicki}},
  \bibinfo{author}{\bibfnamefont{R.}~\bibnamefont{Watkins}}, \bibnamefont{and}
  \bibinfo{author}{\bibfnamefont{K.~A.} \bibnamefont{Olive}},
  \bibinfo{journal}{Nucl.Phys.} \textbf{\bibinfo{volume}{B310}},
  \bibinfo{pages}{693} (\bibinfo{year}{1988}).

\bibitem[{\citenamefont{Edsjo and Gondolo}(1997)}]{Edsjo:1997bg}
\bibinfo{author}{\bibfnamefont{J.}~\bibnamefont{Edsjo}} \bibnamefont{and}
  \bibinfo{author}{\bibfnamefont{P.}~\bibnamefont{Gondolo}},
  \bibinfo{journal}{Phys. Rev. D} \textbf{\bibinfo{volume}{56}},
  \bibinfo{pages}{1879} (\bibinfo{year}{1997}), \eprint{hep-ph/9704361}.

\bibitem[{\citenamefont{Ghosh et~al.}(2022)\citenamefont{Ghosh, Mahapatra,
  Narendra, and Sahu}}]{Ghosh:2021khk}
\bibinfo{author}{\bibfnamefont{P.}~\bibnamefont{Ghosh}},
  \bibinfo{author}{\bibfnamefont{S.}~\bibnamefont{Mahapatra}},
  \bibinfo{author}{\bibfnamefont{N.}~\bibnamefont{Narendra}}, \bibnamefont{and}
  \bibinfo{author}{\bibfnamefont{N.}~\bibnamefont{Sahu}},
  \bibinfo{journal}{Phys. Rev. D} \textbf{\bibinfo{volume}{106}},
  \bibinfo{pages}{015001} (\bibinfo{year}{2022}), \eprint{2107.11951}.

\bibitem[{\citenamefont{Bhattacharya et~al.}(2019)\citenamefont{Bhattacharya,
  Ghosh, and Sahu}}]{Bhattacharya:2018cgx}
\bibinfo{author}{\bibfnamefont{S.}~\bibnamefont{Bhattacharya}},
  \bibinfo{author}{\bibfnamefont{P.}~\bibnamefont{Ghosh}}, \bibnamefont{and}
  \bibinfo{author}{\bibfnamefont{N.}~\bibnamefont{Sahu}},
  \bibinfo{journal}{JHEP} \textbf{\bibinfo{volume}{02}}, \bibinfo{pages}{059}
  (\bibinfo{year}{2019}), \eprint{1809.07474}.

\end{thebibliography}
\end{document}